\documentclass[12pt]{article}
\usepackage[mathscr]{eucal}
\usepackage{epsfig,amsfonts}
\usepackage{amsmath}
\usepackage{amsthm,amssymb}
\usepackage{graphicx}
\usepackage{hhline}
\usepackage{relsize}
\usepackage{cite}
\usepackage{psfrag}
\usepackage{mathrsfs} 
\usepackage{hyperref}
\usepackage{bm}
\usepackage{array}
\usepackage{tikz}

\usepackage{textcomp}
\usepackage{hyperref}
\usepackage{mathrsfs} 
\usepackage{enumitem}
\DeclareMathAlphabet{\mathpzc}{OT1}{pzc}{m}{it}

\makeatletter
\@addtoreset{equation}{section}
\makeatother

\def\bea{\begin{eqnarray}}
\def\eea{\end{eqnarray}}
\def\be{\begin{equation}}
\def\ee{\end{equation}}

\def\be{\begin{equation}}
\def\ee{\end{equation}}
\def\bdm{\begin{displaymath}}
\def\edm{\end{displaymath}}
\def\bea{\begin{eqnarray}}
\def\eea{\end{eqnarray}}

\def\ri{{\rm i}}
\def\half{\textstyle\frac{1}{2}}
\def\XXint#1#2#3{{\setbox0=\hbox{$#1{#2#3}{\int}$}
    \vcenter{\hbox{$#2#3$}}\kern-.5\wd0}}

\newcommand{\rd}{\mbox{d}}
\newcommand{\re}{\mbox{e}}

\DeclareMathAlphabet{\mathpzc}{OT1}{pzc}{m}{it}

\begin{document}

\begin{titlepage}
\begin{flushright}
$\phantom{{\it tresrtfdsgqw }}$\\
\end{flushright}
\begin{flushright}
DESY 19-168 \\
\end{flushright}

\vspace{0.8cm}

\begin{center}
\begin{LARGE}

{\bf Spectrum of  the reflection operators in \\
\vspace{0.2cm}
different integrable structures
}

\end{LARGE}
\vspace{1.3cm}
\begin{large}

{\bf 
%Vladimir V. Bazhanov$^{1}$, 
 Gleb A.  Kotousov$^{1}$   
%\bigskip
%Sergii Koval$^{1}$ 
and 
Sergei  L. Lukyanov$^{2,3,4}$
}

\end{large}

\vspace{1.cm}
$^1$DESY, Theory Group, Notkestrasse 85, Hamburg, 22607, Germany\\
\vspace{.4cm}

%$^2$Mathematical Sciences Institute\\
%      Australian National University, Canberra, ACT 0200,
%      Australia\\\ \\
${}^{2}$NHETC, Department of Physics and Astronomy \\
     Rutgers University\\
     Piscataway, NJ 08855-0849, USA\\
\vspace{.4cm}

${}^{3}$Kharkevich Institute for Information Transmission Problems,\\
Moscow, 127994, Russia

\vspace{.4cm}
and\\

\vspace{.4cm}
${}^{4}$International Institute of Physics,  
Natal-RN 59078-400, Brazil
\vspace{1.0cm}

\end{center}

\begin{center}
\centerline{\bf Abstract} \vspace{.8cm}

\parbox{13cm}{%
The reflection operators are the simplest examples of the 
non-local integrals of motion, which appear in many interesting
problems in integrable CFT.
For the so-called
Fateev, quantum AKNS,
paperclip and KdV integrable structures, they are built from
the (chiral) reflection $S$-matrices for the Liouville and cigar CFTs.
Here we give the full spectrum of the 
reflection operators associated with these integrable structures.
We also obtained a relation between the 
reflection $S$-matrices of the cigar and Liouville CFTs.
The results of this work  are applicable for the description of the scaling behaviour
of the Bethe states in exactly solvable lattice systems and
may be of interest to the study of 
the Generalized Gibbs Ensemble
 associated with the above mentioned integrable structures.

}
\end{center}

\vfill

\end{titlepage}
\setcounter{page}{2}

 \section{Introduction}
 
 The problem of the simultaneous diagonalization of  an infinite set of mutually commuting local Integrals of Motion (IM)
naturally appears in  the study of  1+1 dimensional  integrable QFT.
In the case of a scale invariant theory significant simplifications occur due to the
presence of an infinite dimensional algebra of (extended) conformal
symmetry \cite{Zamolodchikov:1989zs}. 
For a finite-size 2D CFT, where
the spatial coordinate is compactified on a circle, it is possible to give
a mathematically satisfactory construction of an infinite set of
local  IM, whose
simultaneous diagonalization becomes a
well-defined problem within  the representation theory of the associated
conformal algebra.
Different conformal algebras, as well as different sets of mutually
commuting local IM, yield a variety of integrable structures in CFT.

\bigskip
Perhaps the simplest integrable structure involves  the Virasoro algebra itself,
\bea\label{sosososoy}
[L_n,L_m]=(n-m)\, L_{n+m}+\frac{c}{12}\ n(n^2-1)\, \delta_{n+m,0}\,,
%\nonumber
\eea
with the local IM being given by integrals over  the local densities built out of the holomorphic
field
\bea\label{Texpansion1}
 T(u)=-\frac{c}{24}+\sum_{n=-\infty}^\infty L_n\ \re^{-n u}\ \ \ \ \ \ \ (u=x_2+\ri \,x_1)\ .
% \nonumber
\eea
The explicit form for the first few IM reads as follows\footnote{In this work the subscript in the
notation   of the local IM ${\mathbb I}_s$  always 
indicates that  the corresponding local density has the Lorentz spin
$s+1$, e.g., in eq.\,\eqref{KdVint2} 
${\rm spin}(T)=2,\ {\rm spin}(T^2)=4,\ {\rm spin}(T^3)={\rm spin}\big((\partial T)^2\big)=6$.} 
\bea\label{KdVint2}
{\mathbb I}_1^{({\rm KdV})}=\int\limits_0^{2\pi}\frac{\rd x_1}{2\pi}\ T\, ,\ \ 
{\mathbb I}_3^{({\rm KdV})}=\int\limits_0^{2\pi}\frac{\rd x_1}{2\pi}\ T^2
\, ,\ \  {\mathbb I}_5^{({\rm KdV})}=\int\limits_0^{2\pi}\frac{\rd x_1}{2\pi}\ \Big( T^3-\frac{c+2}{12}\, (\partial T)^2\, \Big)\,,
\ldots\ .
\eea
They  are referred to as the local IM for the
quantum KdV integrable structure since,
in the limit when the central charge $c\to\infty$ (which can be understood as a certain classical limit),
the  $\big\{{\mathbb I}_{2m-1}^{({\rm KdV})}\big\}_{m=1}^\infty$ %$\{{\mathbb I}_{2n-1}\}_{n=1}^\infty$
 becomes the set of IM for the classical KdV  equation 
\cite{Sasaki:1987mm,Eguchi:1989hs,Kupershmidt:1989bf, FF5, Bazhanov:1994ft}.
 The operators \eqref{KdVint2} act in the
Verma module ${\cal V}_\Delta$ for the Virasoro algebra. An important property is that they
leave the level subspaces  ${\cal V}^{(N)}_\Delta$ of the 
 Verma module invariant. Thus the problem of their simultaneous diagonalization
can be restricted to ${\cal V}^{(N)}_\Delta$, where the IM are
 ${\tt par}_1(N)\times{\tt par}_1(N)$ dimensional mutually commuting
matrices with ${\tt par}_1(N)$ being the number of integer partitions of $N$.
\bigskip

In ref.\cite{Bazhanov:2003ni} the spectrum of the local IM  \eqref{KdVint2}  in ${\cal V}^{(N)}_\Delta$ was
expressed in terms of the solutions of the algebraic system
 \bea\label{jsaysssysa}
\sum_{b\not=a}^N\frac{v_a\, (\, v_a^2+(3+\alpha)(1+2\alpha) v_a v_b + \alpha(1+2\alpha)\, v_b^2\,)}{(v_a-v_b)^3}
-\frac{ v_a}{4}+\Delta
%\frac{(2\ell+1)^2-4\alpha^2}{16 (\alpha+1)}
=0\ \ \ \ \ \ \ \ (a=1,\ldots, N)\ ,
\eea
where  $\alpha$ parameterizes the central charge
\bea
c=1-\frac{6\alpha^2}{\alpha+1}\ .
\eea 
It was conjectured in that work and recently proved
 by D. Masoero \cite{Masoero}
that for generic (complex) values of $\Delta$ and $c$
the number of distinct, up to the action of the symmetric group $S_N$,
 solutions of the algebraic system \eqref{jsaysssysa} coincides with 
 the number of integer partitions ${\tt par}_1(N)$. The eigenvalues
of the local IM $\mathbb{I}_{2m-1}^{({\rm KdV})}$ turn out to be 
certain symmetric polynomials of order $m-1$ w.r.t. $\{v_a\}_{a=1}^N$
that satisfy the set of equations \eqref{jsaysssysa}.
This implies  that  these  distinct unordered sets  
can be used to label states,
which form  a basis in ${\cal V}^{(N)}_\Delta$ and
 simultaneously diagonalize the  local IM  \eqref{KdVint2}:
 \bea\label{aosaisais}
 {\mathbb I}_{2m-1}^{({\rm KdV})}\,|\,{\boldsymbol v}\,
\rangle={I}_{2m-1}^{({\rm KdV})}({\boldsymbol v})\,|\,{\boldsymbol v}\,\rangle\ ,\ \ \ \ 
 {\rm where}\ \ \ \ \bm{v}=(v_1,\ldots,v_N)\ .
 \eea

\bigskip
Together with the local IM the integrable structures also involve the non-local IM. In the
case of the quantum KdV integrable structure these were  discussed in 
ref.\cite{Bazhanov:1996dr%,Bazhanov:1998dq
}. Furthermore, in ref.\cite{Zamolodchikov:1995aa},
it was pointed out that the set of non-local IM contains an %important 
operator  which is deeply related to the Liouville reflection $S$-matrix.
%In what follows we will refer to this operator as the reflection operator and denote it by ${\mathbb R}$.
Its construction is based on the observation that for $c\ge25$ the field $T(u)$ 
\eqref{Texpansion1} can be interpreted as the holomorphic component of the stress-energy tensor
for the Liouville CFT. The configuration space of the Liouville theory contains a domain where
the exponential interaction term is negligible and the Liouville field becomes a free massless field,
whose chiral component is built out of the 
Heisenberg  operators
\bea\label{hssysy}
[a_n,a_m]=\frac{n}{2}\ \delta_{n+m,0} \ .
\eea
 In the asymptotic domain, 
the Virasoro generators are expressed in terms of the Heisenberg ones as
\bea\label{ososossi}
L_n&=&\sum_{m\not=0,n}a_m a_{n-m}+\big(2 a_0+\ri Q\, n\big)\, a_n\ \ \ \  \ \ \ \ \ (n\not=0)\nonumber\\
L_0&=&2\sum_{m>0}a_{-m} a_{m}+a_0^2+\tfrac{1}{4}\, Q^2\, ,
\eea
where
\bea
Q=\sqrt{\frac{c-1}{6}}
%=\frac{\ri\alpha}{\sqrt{\alpha+1}}
\ .
\eea

The Heisenberg algebra \eqref{hssysy} can be represented in the Fock space ${\cal F}_P$ generated
by the action of the creation operators on the Fock vacuum
\bea\label{vac1a}
a_{n}\,|{ P}\rangle=0\ \ \ \ (\forall \ n>0)\ , \ \ \ \ \ \ \ \ a_0\,|{P}\rangle={P}\,|{ P}\rangle\ .
\eea
As a consequence of the relations \eqref{ososossi},
${\cal F}_P$ is isomorphic to
the Verma module ${\cal V}_\Delta$ with the highest weight $\Delta$ given in terms of
 the zero-mode momentum $P$ as
\bea\label{deltarel1}
\Delta=P^2+\tfrac{1}{4}\,Q^2.
\eea
Notice that $\Delta$ is independent of the sign of $P$.
For $P>0$,  ${\cal F}_P$ and ${\cal F}_{-P}$ can be
interpreted as 
the space of ``in'' and ``out'' 
(chiral) asymptotic states of the Liouville CFT.
It is natural to introduce the $S$-matrix intertwining the two spaces
\bea\label{iaisissai}
{\hat s}_{\rm L}(P\,|\,Q)\ :\ \ \ \ {\cal F}_{P}\mapsto {\cal F}_{-P}\ ,
\eea 
where we explicitly indicate the dependence on the zero mode momentum $P$ and the
 parameter $Q$.
As was discussed in \cite{Zamolodchikov:1995aa}, this operator is
fully determined by the conformal symmetry and the normalization condition
\bea
{\hat s}_{\rm L}(P\,|\,Q)\,|P\rangle=|-P\rangle\ .
\eea
%It should be emphasized that, since this operator intertwines
%different Fock spaces, the problem of its diagonalization does not make sense.
%However 
One can introduce another intertwiner, the ``$C$-conjugation'',  whose action is defined by the relations
\bea\label{oosostr}
{\hat  C}\ :\ \ \ \ {\hat C}\, a_{n}\, {\hat C}=-a_{n}\ ,\  \ \ \ {\hat C}\,|P\rangle=|-P\rangle\ ,
%\nonumber
\eea
so that the operator ${\hat  C}{\hat { s}}_{\rm L}$  acts invariantly in the Fock space.
It turns out that the latter commutes with the local IM \eqref{KdVint2} provided
that they are understood as operators in ${\cal F}_P$ via eq.\,\eqref{ososossi}
\cite{Zamolodchikov:1995aa,Kotousov:2019ygw}.
% and
%\bea\label{owieoi223}
%{\hat  C}{\hat { s}}(p)\, :\ \ \ \  {\cal F}_{\tt p}\mapsto  {\cal F}_{\tt p}\ 
%\ \ \ \ \ {\rm and} \ \ \ \ \ \   {\hat  C}{\hat { s}}(p)\,|p\rangle=|p\rangle\ .
%\eea
%commutes with the action of the local IM \eqref{KdVint2} .
We define the reflection operator associated with the quantum KdV integrable structure as
in  ref.\!\cite{Kotousov:2019ygw}
\bea\label{sosoisis}
{\mathbb R}^{({\rm KdV})}=R_{0}\ \big[{\hat  C}{\hat s}_{\rm L}(P|\,Q)\big]^{-1}\, .
\eea
The scalar factor $R_{0}$ coincides with the vacuum  eigenvalue and its value is
 not essential for the purpose of this work.
Thus we arrive at the problem
of the calculation of the eigenvalues $R^{({\rm KdV})}({\boldsymbol v})$:
\bea\label{hassayyas}
{\mathbb R}^{({\rm KdV})}\,|{\boldsymbol v}\rangle_P=R^{({\rm KdV})}({\boldsymbol v})\, |{ \boldsymbol v}\rangle_P\ ,
\eea
where  we emphasize with the subscript ``$P$''  that
$|{ \boldsymbol v}\rangle_P$ should be considered
as a state in the
 level subspace ${\cal F}_P^{(N)}\cong{\cal V}^{(N)}_\Delta$.
Among the purposes of this work is to present an explicit description of
$R^{({\rm KdV})}({\boldsymbol v})$ in terms of the  
sets $\{v_a\}_{a=1}^N$ solving eqs.\eqref{jsaysssysa}.

\bigskip
In the case of extended conformal symmetry the r$\hat{{\rm o}}$le of the symmetry algebra is
usually played  by a $W$-algebra
generated by the holomorphic fields of (half-) integer Lorentz spin along with the spin-2 field similar to
$T(u)$ \eqref{Texpansion1} \cite{Zamolodchikov:1985wn}. It should be kept in mind that  for a given such algebra
there are a number of integrable structures corresponding to the different
sets of mutually commuting  local IM that can be built out of the holomorphic fields.
 For example, in the case of the $W_\infty$-algebra studied  in \cite{Bakas:1991fs},
there are at least three different integrable structures: the quantum AKNS \cite{Fateev:1995ht,
Fateev:2005kx,Bazhanov:2008yc,Bazhanov:2019xvy}, 
paperclip
 \cite{Fateev:1995ht,Lukyanov:2003nj,Bazhanov:2017nzh} and
Getmanov ones \cite{Fateev:1990bf,Lukya}. 
For each of them there is an associated  reflection operator(s)
similar to   \eqref{sosoisis}.
In this work we consider the eigenvalue problem
 for  the reflection operators for  the 
quantum AKNS and paperclip integrable structures.
It turns out that the spectrum of these reflection operators as well as ${\mathbb R}^{({\rm KdV})}$
\eqref{sosoisis} can be obtained via certain limiting procedures and restrictions from
that of the reflection operators introduced  in ref.\cite{Bazhanov:2013cua}.  
The latter appear within  the integrable structure
originally found  by Fateev \cite{Fateev:1995ht,Fateev:1996ea,Lukyanov:2012wq}.
%and studied in \cite{Lukyanov:2012wq}
% in the context of the so-called pillow-brane model.

 \bigskip
 The paper is organized as follows. In sec.\,\ref{stwo} we use the results of the recent work
 of Eremenko and Tarasov \cite{ET} to calculate the spectrum of the reflection operators in the
 Fateev integrable structure. As will be discussed in the subsequent section, this
  allows one to  find the spectrum of the reflection operators in the quantum AKNS integrable structure
 through  a certain limiting procedure. In sec.\,\ref{sthree} we also list some basic facts about
 the quantum AKNS and Getmanov integrable structures.
 Sec.\,\ref{sfour} is devoted to the
 reflection operator for the paperclip integrable structure. 
Its spectrum follows immediately from the results of sec.\,\ref{stwo} 
by a specialization of the parameters.
In turn,  a proper restriction of the  formulae in sec.\,\ref{sthree}
yields the spectrum of the reflection operator \eqref{sosoisis}.
 This is the subject of consideration of 
 sec.\,\ref{sfive}. 
Finally, the relation of the reflection operators
to the different  Hermitian forms associated with the  quantum AKNS and paperclip integrable structures
is briefly discussed in sec.\,\ref{ssix}.

 \section{\label{stwo}Spectrum of the reflection operators in the Fateev integrable structure}

The $W$-algebra associated with the exceptional 
 Lie superalgebra $D(2,1;\alpha)$ that was studied  in ref.\cite{Feigin:2001yq}
appeared in the description of the so-called corner-brane  conformal boundary state in  \cite{Lukyanov:2012wq},
 where it was referred to as the ``corner-brane $W$-algebra''.
 This algebra of extended conformal symmetry admits the integrable structure
 that was found in  \cite{Fateev:1996ea} and further investigated in
the papers \cite{Lukyanov:2012wq,Bazhanov:2013cua}.
 In particular, the construction of the corresponding reflection operators
  was given in sec.\,(6.1) from \cite{Bazhanov:2013cua}.
  % where
 %they were denoted by ${\mathbb R}^{(k)}_{\sigma'\sigma}$ with $(k=1,2,3; \ \sigma',\sigma=\pm 1)$.
 The reflection operators act invariantly in the level subspaces
 of  the Fock space  which is the highest weight representation
 of three copies of the Heisenberg algebra of the form
  \eqref{hssysy}, labeled by the three component zero-mode momentum.
 %The eigenvalues of this reflection operators in the level  ${\cal F}^{(L)}_{\bf P}$ take the form
 %\bea
 %{\mathbb R}^{(k)}_{\sigma'\sigma}\,|{\boldsymbol x}\rangle=\big(g^{(L)}_j({\boldsymbol x})\big)^{\sigma'}
 %\big(g^{(L)}_i({\boldsymbol x})\big)^{\sigma} \,|{\boldsymbol x}\rangle
% \eea
 %Here $(i, j, k)$ is any cyclic permutation of $(1, 2, 3)$, $\sigma,\sigma'=\pm 1$ and
 %${\boldsymbol x}$ stands for the non-ordered set $\{x_a\}_{a=1}^L$, satisfying a certain
 %algebraic system which is somewhat similar  \eqref{jsaysssysa}  $\{v_a\}_{a=1}^L$  \eqref{jsaysssysa}.
 %In  the work \cite{Bazhanov:2013cua}  
 The problem
 of the calculation of the spectrum of the reflection operators was reduced to finding 
 connection coefficients  for a certain linear ODE. Let us first give an outline of the relevant results
 from ref.\cite{Bazhanov:2013cua}.

 \bigskip
 The starting point of  that work  is the so-called generalized hypergeometric equation
\bea\label{aoossaisa}
\big(-\partial_z^2+t(z)\big)\, \psi=0\ ,
\eea
where
\bea\label{sospsaopsa}
t(z)=t_0(z)+t_1(z)
\eea
with
\bea\label{osaspsosap}
t_0(z)&=&-\sum_{i=1}^3\Big(\frac{\delta_i}{(z-z_i)^2}+\frac{c_i}{z-z_i}\,\Big) \\
t_1(z)&=&\sum_{a=1}^L\Big(\frac{2}{(z-x_a)^2}-\frac{C_a}{z-x_a}\,\Big)\ .\nonumber
\eea
Treating the complex variable $z$ as a coordinate on the Riemann sphere, the imposed conditions
\bea\label{usosopsaopspoa}
&&\sum_{i=1}^3c_i=-\sum_{a=1}^L C_a\ \nonumber\\
&&\sum_{i=1}^3(c_i\, z_i+\delta_i)=\sum_{a=1}^L\big(2-C_a x_a\big)\\
&&\sum_{i=1}^3(c_i\, z^2_i+2\delta_i\, z_i)=\sum_{a=1}^L\big(4 x_a-C_a\, x^2_a\big)\nonumber
\eea
imply that the north pole of the sphere corresponding to  $z=\infty$ is a regular point for the ODE
\eqref{aoossaisa}.
Equations \eqref{usosopsaopspoa} can be used to express $c_1,c_2,c_3$ through the other
parameters.
 Moreover it is assumed that 
$\{x_a\}_{a=1}^L$ and $\{C_a\}_{a=1}^L$
obey the system of algebraic equations
\bea\label{spossaopa}
&&C_a\ \bigg[\ \frac{1}{4}\,
C_a^2-t_0(x_a)-\sum_{b\not= a}^L\Big(\frac{2}{(x_a-x_b)^2}-\frac{C_b}{x_a-x_b}\,\Big)
\bigg]\nonumber\\
&&-\,
t'_0(x_a)+\sum_{b\not= a}^L\bigg(\frac{4}{(x_a-x_b)^3}-\frac{C_b}{(x_a-x_b)^2}\,\bigg)
=0\ \ \ \ \ \ \ (a=1,\ldots,L)\ ,
\eea
where the prime stands for the derivative w.r.t. the argument.
The above system  guarantees   that all the singularities at $z=x_a\ (a=1,\ldots,L)$ are apparent, i.e.,
 any solution of  \eqref{aoossaisa} remains a single valued function in the
 vicinity of these singular points.
\bigskip
 
Let $\chi^{(i)}_\sigma(z)$ $(i=1,2,3;\ \sigma=\pm)$  be  solutions  of \eqref{aoossaisa}
such that
\bea\label{opaosap}
\chi_\sigma^{(i)}\to\frac{1}{\sqrt{2p_i}}\  (z-z_i)^{\frac{1}{2}+\sigma
  p_i}\,\Big(1+O(z-z_i)\Big)\qquad {\rm as}\ \ \ \ \ \ \ \ \ z\to z_i
\eea
with  the $p_i$ defined by the relations
\bea
\delta_i=\tfrac{1}{4}-p_i^2\,.
\eea
The prefactor in eq.\,\eqref{opaosap}  is chosen to ensure the following normalization for $\chi_\sigma^{(i)}$:
\bea\label{aoposas}
{\tt W}[\chi_{\sigma'}^{(i)},\,\chi_\sigma^{(i)}]=\sigma\,  \delta_{\sigma+\sigma',0}\ ,
\eea
where ${\tt W}[f,g]=f g'-g  f'$ stands for the Wronskian.
With the restriction $0<p_i<\frac{1}{2}$, the asymptotic conditions \eqref{opaosap}
define three
different bases (for $i=1,2,3$) in the  two-dimensional  
linear space of solutions of \eqref{aoossaisa}.
Let us  combine the solutions for given $i$ into the row vector
\bea\label{chi-basis}
{\boldsymbol \chi}^{(i)}=(\chi_-^{(i)},\chi_+^{(i)})\,,\qquad i=1,2,3\ .
\eea
Then the linear transformation relating any two sets of bases can be
expressed in the form
\bea\label{sksssopsosp}
{\boldsymbol \chi}^{(i)}={\boldsymbol \chi}^{(j)}
\ {\boldsymbol S}^{(j,i)}\ .
\eea
The matrices 
 \bea
 {\boldsymbol S}^{(j,i)}=\begin{pmatrix}
S^{(j,i)}_{--}&S^{(j,i)}_{-+}\\
S^{(j,i)}_{+-}&S^{(j,i)}_{++}
\end{pmatrix}
\eea
obey the relations
\bea\label{saopsapo}
\det\big({\boldsymbol S}^{(j,i)}\big)=1\,,\qquad
{\boldsymbol S}^{(i,j)}{\boldsymbol S}^{(j,i)}={\boldsymbol I}\,,\qquad
{\boldsymbol S}^{(i,k)}{\boldsymbol S}^{(k,j)}{\boldsymbol S}^{(j,i)}={\boldsymbol I}\,,
\eea
where in the last equality $(i,j,k)$ is any cyclic permutation of $(1,2,3)$.
In the case $L=0$,  the solutions of  the ODE \eqref{aoossaisa} are given in terms of the
hypergeometric function and the matrices ${\boldsymbol S}^{(j,i)}|_{L=0}$ 
are well known  (see, e.g., eqs.(2.35)-(2.37)  and B.2 in \cite{Bazhanov:2013cua}). 
For general $L=1,2,\ldots$ they can be represented in the form
\bea\label{Gfunc1}
{ S}^{(2,1)}_{\sigma'\sigma}=G_L({\boldsymbol X}|
-\sigma p_1,\sigma'p_2,p_3)\ { S}^{(2,1)}_{\sigma'\sigma}\big|_{L=0}\  \ \ \ \ \ \ \ \ \  (\sigma,\sigma'=\pm)\,,
\eea
where the function $G_L$ depends on the 
$L$ cross-ratios ${\boldsymbol X}=(X_1,X_2,\ldots,X_L)$,
 \bea\label{aoisisai}
 X_a=\frac{x_a-z_1}{x_a-z_3}\ \frac{z_2-z_3}{z_2-z_1}\ ,
 \eea
as well as the parameters  $(p_1,p_2,p_3)$.
 
The ODE 
 \eqref{aoossaisa}-\eqref{spossaopa} is covariant w.r.t. M${\ddot {\rm o}}$bius  transformations of the Riemann
 sphere. This can be used to move the non-apparent  singularities to the standard positions:
 \bea\label{aiaiasis}
 (z_1,z_2,z_3)=(0,1,\infty)\ .
 \eea
 With this set up the cross-ratios $X_a$ \eqref{aoisisai} coincide with $x_a$ and
 \bea\label{sooad}
t_0(z)&=&
-\,\bigg[\ \frac{\delta_1}{z^2}+\frac{\delta_1+\delta_2-\delta_3-2L-\sum_{a=1}^LC_a\, (1-x_a)}{z}\nonumber\\
&+&
\frac{\delta_2}{(z-1)^2}-\frac{\delta_1+\delta_2-\delta_3-2L+\sum_{a=1}^LC_a\, x_a}{z-1}\ \bigg]\\[0.2cm]
t_1(z)&=&\sum_{a=1}^L\bigg(\frac{2}{(z-x_a)^2}-\frac{C_a}{z-x_a}\,\bigg)\ .\nonumber
\eea
In what follows we will always assume the choice \eqref{aiaiasis}, so that
 \bea\label{aossaiasi}
 G_L=G_L({\boldsymbol x}\,|\,p_1,p_2,p_3)\ ,\ \ \ \ \ \ {\rm where}\ \ \ \ {\boldsymbol x}=\bm{X}=(x_1,x_2,\ldots,x_L)\ .
 \eea

 \bigskip
 The set of $L$ equations \eqref{spossaopa} with $t_0(z),\,t_1(z)$ given by \eqref{sooad} allows one to express
 $\{C_a\}_{a=1}^L$ in terms of the $
 \{x_a\}_{a=1}^L$. In ref.\!\cite{Bazhanov:2013cua} it was conjectured that if this algebraic system  is supplemented 
 by the extra equations
 \bea\label{aiisiss}
 C_a=\frac{2-a_1}{x_a}+\frac{2-a_2}{x_a-1}\ \ \ \ \ \ \ \ \ (a=1,\ldots,L)\ ,
 \eea
 then for generic values of the parameters $a_1,\ a_2$ and $(p_1,p_2,p_3)$, the combined system
 admits exactly ${\tt par}_3(L)$ distinct  (up to  permutations) solutions for $\{x_a\}_{a=1}^L$.
 Here ${\tt par}_3(L)$ denotes the number of partitions of  $L$ into integer parts of three kinds:
 \bea
 \sum_{L=0}^\infty{\tt par}_3(L) \ q^L=\prod_{m=1}^\infty\frac{1}{(1-q^m)^3}=1+3\,q+9\,q^2+22\, q^3+\ldots\ .
 \eea
The sets of solutions ${\boldsymbol x}=(x_1,\ldots,x_L)$  label the  basis states in
 the level-$L$  subspaces  of the Fock space whose highest weight is parameterized by ${\boldsymbol p}=(p_1,p_2,p_3)$.
This particular basis diagonalizes the local IM from the Fateev integrable structure:
 \bea
 {\mathbb I}^{(\rm F)}_{2m-1}\, |{\boldsymbol x}\rangle_{\bm p}
 = I^{(\rm F)}_{2m-1}({\boldsymbol x})\,  |{\boldsymbol x}\rangle_{{\bm p}}
 \eea
and the eigenvalues $I^{(\rm F)}_{2m-1}({\boldsymbol x})$ turn out to be symmetric functions of $x_a$
(for an  illustration see eqs.(3.40)-(3.44), (4.17) from ref.\cite{Bazhanov:2013cua}).
 Notice  that
 these  local IM %in the Fateev  integrable structure, $\{{\mathbb I}^{(F)}_{2n-1}\}_{n=1}^\infty$,
 depend additionally
 on the two parameters $a_1$ and $a_2$ which appear  in eq.\eqref{aiisiss}.
\bigskip

The reflection operators for the Fateev integrable structure
were introduced in the work \cite{Bazhanov:2013cua}.
The paper considered twelve operators, 
denoted as  ${\mathbb R}^{(k)}_{\sigma'\sigma}\ (k=1,2,3;\,\sigma,\sigma'=\pm)$,
such that
\be
{\mathbb R}^{(k)}_{-\sigma'-\sigma}=\big[{\mathbb R}^{(k)}_{\sigma'\sigma}\big]^{-1}\ ,
\qquad \quad \big[\,{\mathbb R}^{(k)}_{\sigma'\sigma}, {\mathbb I}^{(\rm F)}_{2m-1}\,\big]=0\ .
\ee
For their explicit construction, we refer the reader to sec.\,(6.1) in \cite{Bazhanov:2013cua}.
The results of that work imply that the
eigenvalues of ${\mathbb R}^{(3)}_{\sigma'\sigma}$ for the state $|{\boldsymbol x}\rangle_{{\bm p}}$ are given by
 \bea\label{kasissauas}
 R^{(3)}_{++}&=&\big[R^{(3)}_{--}\big]^{-1}=G_L({\boldsymbol x}\,|\,p_1,p_2,p_3\!\parallel\!a_1,a_2)
\nonumber\\[-0.2cm]
&&\\[-0.2cm]
 R^{(3)}_{-+}&=&\big[R^{(3)}_{+-}\big]^{-1}=G_L({\boldsymbol x}\,|\,p_1,-p_2,p_3\!\parallel\!a_1,a_2) \ .
\nonumber
\eea
Note that, having made the specialization \eqref{aiisiss},
the function 
\be\label{Gfunc1a}
 G_L=G_L({\boldsymbol x}\,|p_1,p_2,p_3\!\parallel\!a_1,a_2)
\ee
depends additionally on the two parameters $a_1$ and $a_2$
as well as $\bm{x}=(x_1\,\ldots,x_L)$, which is now understood as one of the ${\tt par}_3(L)$  solutions of the system
\eqref{spossaopa},\eqref{sooad} and \eqref{aiisiss}.
\bigskip

It is possible to obtain the eigenvalues of the
 reflection operators  ${\mathbb R}^{(k)}_{\sigma'\sigma}$ with $k=1,2$ through the modular transformation
of the set $\bm{x}$.
In particular, one has
\bea
 R^{(2)}_{++}&=&\big[R^{(2)}_{--}\big]^{-1}=G_L(\tilde{\boldsymbol x}\,|\,p_3,p_1,p_2\!\parallel\!a_3,a_1)\nonumber\\[-0.0cm]
&&\\[-0.0cm]
 R^{(2)}_{-+}&=&\big[R^{(2)}_{+-}\big]^{-1}=G_L(\tilde{\boldsymbol x}\,|\,p_3,-p_1,p_2\!\parallel\!a_3,a_1) \nonumber
\eea
and
\bea
 R^{(1)}_{++}&=&\big[R^{(1)}_{--}\big]^{-1}=G_L(\tilde{\tilde{\boldsymbol x}}\,|\,p_2,p_3,p_1\!\parallel\!a_2,a_3) \nonumber\\[-0.0cm]
&&\\[-0.0cm]
 R^{(1)}_{-+}&=&\big[R^{(1)}_{+-}\big]^{-1}=G_L(\tilde{\tilde{\boldsymbol x}}\,|\,p_2,-p_3,p_1\!\parallel\!a_2,a_3)\ ,\nonumber
 \eea
where
$$
\tilde{{\boldsymbol x}}=\big(\tfrac{1}{1-x_1},\,\ldots,\tfrac{1}{1-x_L}\big)\, ,\qquad 
\quad \tilde{\tilde{\boldsymbol x}}=\big(1-\tfrac{1}{x_1},\ldots,1-\tfrac{1}{x_L}\big)\ .$$
Also we use the parameter $a_3$ defined by
\be\label{ferkaskd12}
a_1+a_2+a_3=2\ .
\ee 
Thus the problem of calculation of the spectrum of the reflection operators in the Fateev 
 integrable structure is reduced to finding the functions $G_L$ \eqref{Gfunc1}.
 For $L=1$ it was solved in Appendix A of ref.\!\cite{Bazhanov:2013cua}. The results of the
 important work \cite{ET} allows one to  derive an explicit analytical expression for 
 $G_L$ for any $L=1,2,\ldots$\ .
 \bigskip
 \bigskip

The ODE studied by
 Eremenko and Tarasov  
is more general than the one defined by eqs.\,\eqref{aoossaisa},\,\eqref{sooad} and \eqref{spossaopa}. 
Their result, specialized to
 the case  considered here, implies that the basic solutions $\chi_\sigma^{(i)}$
are given  in terms of the conventional hypergeometric function ${}_2F_1$. 
 In particular, assuming that
 the non-apparent singularities are chosen as in \eqref{aiaiasis}, one has
 \bea\label{sisiaasiias}
\chi^{(1)}_-&=&\frac{z^{\frac{1}{2}-p_1}\ (1-z)^{\frac{1}{2}-p_2}}{ \sqrt{2p_1}\,P_{2L}(0)
\prod_{a=1}^L (1-\frac{z}{x_a})}\nonumber\\[0.3cm]
&\times&
P_{2L}\big(z\,\tfrac{\rm d }{{\rm d} z}\big)\, 
{}_2F_1\big(\tfrac{1}{2}-L-p_1-p_2-p_3,\tfrac{1}{2}-L-p_1-p_2+p_3 ,1-2p_1;z\big)\nonumber\\[0.3cm]
 && \\[-0.2cm]
\chi^{(1)}_+&=&\frac{z^{\frac{1}{2}-p_1}\ (1-z)^{\frac{1}{2}-p_2}}{\sqrt{2p_1}\, P_{2L}(2p_1)\prod_{a=1}^L(1-\frac{z}{x_a})}\nonumber\\[0.3cm]
&\times& 
P_{2L}\big(z\,\tfrac{{\rm d } }{{\rm d} z}\big)\, z^{2p_1}\,
{}_2F_1\big(\tfrac{1}{2}-L+p_1-p_2-p_3,\tfrac{1}{2}-L+p_1-p_2+p_3 ,1+2p_1;z\big)\ .\nonumber
\eea
 Here $P_{2L}(D)$ is a certain polynomial  of order $2L$ 
 in the variable $D$ such that
$P_{2L}(D)=\prod_{b>a}(x_b-x_a)\ D^{2L}+\ldots$\ .
It is given by the determinant of an $L\times L$  matrix

  \bea
  P_{2L}(D)=\det\big(x^{b-1}_a\,U_a(D+b)\big)\ \ \ \ \ \  (a,b=1,\ldots, L)\ ,
  \eea
 where $U_a(D) $ entering into the above formula are rather cumbersome functions:
 \bea\label{aoisisaias}
&&U_a(D)
=  (D-1)^2-
 \bigg( \frac{x_a}{x_a-1}\ (2p_2+1) +2p_1+2-C_ax_a+
\sum_{b\not= a}\frac{4 x_a}{x_a-x_b}\bigg)\ (D-1)\nonumber\\
&&+\,\frac{x_a^2}{2}\  C^2_a
+
 \Big( (p_1+p_2)(p_1+p_2+1)-p_3^2+\tfrac{1}{4}+2L+
\big(\, p_1+\tfrac{3}{2}-
 (p_1+p_2+3)\,x_a \, \big)\,C_a\, 
\Big)\ \frac{x_a}{x_a-1}\nonumber\\[0.2cm]
&&+\,2p_1+1+\frac{x^2_a}{(x_a-1)^2} \
(2p_2+1)+\bigg(\sum_{b\not=a}\frac{2x_a}{x_a-x_b}\,\bigg)^2
\\[0.2cm]
&&+\, \sum_{b\not=a}^L\bigg[\,\frac{x_ax_b (x_b-1)\,C_b}{(x_a-x_b)(x_a-1)}
+\bigg(
\frac{x_a(2p_2+1)}{x_a-1}+2p_1+1-C_a x_a \bigg)\ \frac{2x_a}{x_a-x_b}\,\bigg]\ .\nonumber
\eea
 With this result at hand, it is straightforward to show that $G_L$ \eqref{Gfunc1a}
 is given by
 \bea\label{isisisai}
\!\!\! &&G_L({\boldsymbol x}\,|\,p_1,p_2,p_3\!\parallel\!a_1,a_2)
= \\[0.2cm]
&&\frac{\prod_{b>a}(x_b-x_a)}{ \det\big(x^{b-1}_a\,U_a(b)\big)}\, \prod_{b=1}^L \frac{x_b}{x_b-1}
\big( p_1 +p_2 +p_3+b-\tfrac{1}{2}\big) \big( p_1 +p_2 -p_3+b-\tfrac{1}{2}\big)\,\bigg|_{C_a=\frac{2-a_1}{x_a}+\frac{2-a_2}{x_a-1}} .\nonumber
 %\\
% \frac{\Gamma(\frac{1}{2}  - p_1 -p_2 -p_3)\, \Gamma(\frac{1}{2}  - p_1 - p_2 + p_3)}
%{\Gamma(\frac{1}{2}-L  - p_1 -p_2 -p_3)\, \Gamma(\frac{1}{2}-L  - p_1 - p_2 + p_3)}\ .
\eea
 Thus formulae 
 \eqref{kasissauas}-\eqref{ferkaskd12},\,\eqref{isisisai},\,\eqref{aoisisaias} provide a full description
 of the spectrum of the reflection operators  ${\mathbb R}^{(k)}_{\sigma'\sigma}$.
 Recall that  ${\boldsymbol x}=(x_1,\ldots, x_L)$ solves   the algebraic system of equations
 \eqref{spossaopa},\,\eqref{sooad},\,\eqref{aiisiss}. 
%Of course, similar formulae hold true
%  for the eigenvalues of the reflection operators ${\mathbb R}^{(k)}_{\sigma'\sigma}$ with $k=1,2$.
 Needless to say that for the case $L=1$  the general formula for the spectrum of the
 reflection operator turns out to be  equivalent to the result obtained in ref.\cite{Bazhanov:2013cua}.

 \section{\label{sthree}Quantum AKNS integrable structure}
\subsection{\label{sthreeone}Spectrum of the reflection operators}
 As it was pointed out in ref.\cite{Bazhanov:2019xvy}, the quantum  AKNS integrable structure \cite{Fateev:2005kx,Bazhanov:2008yc}  possesses two reflection operators $\check{\mathbb R}$ and  $\check{\mathbb D}$ that commute with
the set of local IM $\big\{{\mathbb I}^{(\rm AKNS)}_{m}\big\}_{m=1}^{\infty}$.
In that work, the eigenvalues of the reflection operators were expressed in terms of the connection coefficients of
the ODE, that can be obtained from \eqref{aoossaisa} 
with $t(z)$ given by eqs.\,\eqref{sospsaopsa},\eqref{sooad},\eqref{spossaopa} through a certain limiting procedure.
The latter is similar to that which brings the Gauss hypergeometric equation to the 
confluent one. For this reason we will refer to the limit as the confluent limit.
\bigskip

 Consider the ODE \eqref{aoossaisa} with $t(z)$ from \eqref{sospsaopsa},\eqref{sooad},\eqref{spossaopa}
such that
\bea
z=\varepsilon\, w\ ,\ \ x_a=\varepsilon\, w_a\ ,\ \ \ C_a=-\varepsilon^{-1}\ \frac{n_a}{w_a}\ , \ \ \ p_1=p,\ \ \   p_2=\tfrac{1}{2}\, \ri s+
\varepsilon^{-1}\ ,\ \ \ 
p_3=\tfrac{1}{2}\, \ri s-
\varepsilon^{-1}
\eea
and $\varepsilon\to 0$. A straightforward calculation leads to the equation
\bea\label{ODE2}
\bigg[-\frac{\rd^2}{\rd w^2}+\frac{p^2-\frac{1}{4}}{w^2}+\frac{2\ri s}{w}+1+\sum_{a=1}^L\bigg(
\frac{2}{(w-w_a)^2}+\frac{n_a}{w(w-w_a)}\bigg)\,\bigg]\ \Psi=0\ .
\eea
Together with a regular singular point at $w=0$, this ODE possesses an irregular singular point at  $w=\infty$, 
as well as $L$ additional  apparent singularities characterized by the $2L$ complex parameters
$(w_a,n_a)$. The latter satisfy    the following system of algebraic constraints
\bea\label{aosaasi}
n_a\, \big(\, \tfrac{1}{4}\, n_a^2- w_a^2\,t_0^{(a)}\big)+w^3_a\, t_1^{(a)}=0 \qquad \qquad (a=1,2,\ldots,L)\,,
\eea
where 
\bea\label{issausau}
t_0^{(a)}&=&\frac{p^2-\frac{1}{4}}{w_a^2}+\frac{2\ri s}{w_a}+1
-\frac{n_a}{w_a^2}+ \sum_{b\not=a}\bigg(\frac{2}{(w_a-w_b)^2}+\frac{n_b}{w_a(w_a-w_b)}\bigg)\\[0.1cm]
t_1^{(a)}&=&-2\ \frac{p^2-\frac{1}{4}}{w_a^3}-\frac{2\ri s}{w_a^2}
+\frac{n_a}{w^3_a}- \sum_{b\not=a}\bigg(\frac{4}{(w_a-w_b)^3}+\frac{n_b\,(2w_a-w_b)}{w_a^2\,(w_a-w_b)^2}\bigg)\ ,\nonumber
\eea
which is just the limiting form of eq.\,\eqref{spossaopa}.
This way we arrive at the ODE appearing in ref.\cite{Bazhanov:2019xvy}. 
Introduce two
 solutions of \eqref{ODE2}-\eqref{issausau} by means of the asymptotic condition
\bea
\Psi_{\pm p}\to w^{\frac{1}{2}\pm p}\ \ \ \ \  \ {\rm with}\ \ \ \ w\to 0\ ,
\eea
and define the connection coefficients $C_p^{(\pm,L)}$ as in eq.(38)  from  \cite{Bazhanov:2019xvy}, i.e.,
\bea
\Psi_p(w)\to
\begin{cases}
C_p^{(+,L)}\ (+w)^{+\ri s}\ \re^{+w}\ \ \ \ {\rm as}\ \ \ \Re e(w)\to+\infty\\
C_p^{(-,L)}\ (-w)^{-\ri s}\ \re^{-w}\ \ \ \ {\rm as}\ \ \ \Re e(w)\to-\infty
\end{cases}\ .
\eea
Taking  the confluent limit of the solutions \eqref{sisiaasiias} 
 one can show that  the connection coefficients $C_p^{(\pm,L)}$ are given by the following formulae
 \bea\label{hsaysayt}
C_p^{(\pm,L)}=C_p^{(\pm,0)}\ \ \frac{(\mp 1)^L\ \det\big(w_a^{b-1}\,{\tilde U}^{(\pm)}_a(b)\big)}
{\prod_{a=1}^L w_a\ \prod_{b>a}(w_b-w_a)\ \prod_{a=1}^L\big(2p+2a-1\pm 2 \ri s\big)}\ ,
\eea
where
\bea\label{aoisisaiaskjj}
{\tilde U}^{(\pm)}_a(D)
&=&  (D-1)^2-
 \bigg( 2p+2+n_a\mp 2w_a+
\sum_{b\not= a}^L\frac{4 w_a}{w_a-w_b}\bigg)\ (D-1)\nonumber\\
&+&\tfrac{1}{2}\, n_a^2+\big(p+\tfrac{3}{2}\big)\ n_a\mp (n_a+1+2p+2\ri s)\ w_a+2 p+1
\\[0.2cm]
&+&\bigg(\sum_{b\not=a}^L\frac{2w_a}{w_a-w_b}\,\bigg)^2
+\, \sum_{b\not=a}^L\big(\,
2 \,(2p+1+n_a\mp 2w_a)-n_b \,\big)\ \frac{w_a}{w_a-w_b}\nonumber
\eea
and
\bea
C_p^{(\pm,0)}=2^{\pm\ri s-p-\frac{1}{2}}\ \frac{\Gamma(1+2p)}{\Gamma(\frac{1}{2}+p\pm\ri s)}\ .
\eea

%\subsection{Spectrum of the reflection operators in the quantum AKNS integrable structure}

The
reflection operators $\check{\mathbb R}^{({\rm AKNS})}$ and  $\check{\mathbb D}^{({\rm AKNS})}$
from the quantum AKNS integrable
structure were normalized in ref.\cite{Bazhanov:2019xvy}
in such a way that their vacuum  (level zero) 
eigenvalues are equal to 1.\footnote{In what follows we will always use the ``check'' notation
for the reflection operators associated with the different integrable structures in
order to emphasize that they are normalized such that their vacuum eigenvalues are equal to one.}
 Then the eigenvalues 
 at level $L$ are expressed through the connection
coefficients $C_p^{(\pm,L)}$  as follows
\bea\label{RDeig1a}
&&\check{R}^{({\rm AKNS})}=\frac{C_p^{(+,L)}}{C_p^{(+,0)}}\ \frac{C_p^{(-,L)}}{C_p^{(-,0)}}\bigg|_{n_a=n}\nonumber\\[-0.3cm]
&&\\
&&\check{D}^{({\rm AKNS})}=\frac{C_p^{(+,L)}}{C_p^{(+,0)}}\ \frac{C_p^{(-,0)}}{C_p^{(-,L)}}\bigg|_{n_a=n}\ .
\nonumber
\eea
Combining these with \eqref{hsaysayt},\,\eqref{aoisisaiaskjj}  and \eqref{aosaasi},\,\eqref{issausau}
 one obtains
\bea\label{haysayas}
\check{R}^{({\rm AKNS})}({\bm w})&=&
 \frac{(-1)^L}{\prod_{a=1}^Lw_a^2}\ 
\frac{\det\big(w_a^{b-1}\,  V^{(+)}_a(b) \big)\,
\det\big(w_a^{b-1}\,V^{(-)}_a(b) \big)}
{\prod_{b>a}(w_b-w_a)^2 \prod_{a=1}^L(2p+2a-1+2\ri s\big)
\big(2p+2a-1- 2\ri s\big)}\nonumber\\[0.2cm]
 \check{D}^{({\rm AKNS})}({\bm w})&=&
(-1)^L\ 
\prod_{a=1}^L\frac{p+a-\frac{1}{2}-\ri s}{p+a-\frac{1}{2}+\ri s}\ \ \ 
\frac{\det\big(w_a^{b-1}\, V^{(+)}_a(b)\big)}{\det\big(w_a^{b-1}\,V^{(-)}_a(b)\big)}\ .
 \eea
 Here
\bea\label{aoiaiaskjjsss}
V^{(\pm)}_a(D)
&=&  (D-1)^2-
 \bigg( 2p+2+n\mp 2w_a+
\sum_{b\not= a}^L\frac{4 w_a}{w_a-w_b}\bigg)\ (D-1)\nonumber\\
&+&\tfrac{1}{2}\, n^2+\big(p+\tfrac{3}{2}\big)\ n\mp (n+1+2p\pm 2\ri s)\ w_a+2 p+1
\\[0.2cm]
&+&\bigg(\sum_{b\not=a}^L\frac{2w_a}{w_a-w_b}\,\bigg)^2
+\, \big(\,
4p+2\mp 4w_a+n \,\big)\,
\sum_{b\not=a}^L\frac{w_a}{w_a-w_b}\nonumber
\eea
and the set $\bm{w}=\{w_a\}_{a=1}^L$ obeys the system of algebraic equations
\bea\label{sksksk1}
4 n\, w_a^2\!\!&+&\!\!8\ri s\, (n+1)\, w_a-(n+2)\ \big((n+1)^2-4p^2\big)\\[0.2cm]
&+&\!\!
4\ \sum_{b\not=a}^L\frac{w_a\, (\, (n+2)^2\, w_a^2- n(2n+5)\, w_a w_b + n(n+1)\, w_b^2\,)}{(w_a-w_b)^3}=0\  \ \ \ \ \ \ \ \ (a=1,\ldots,L)\ .\nonumber
\eea
Note that for $L=1$ eqs.\,\eqref{haysayas},\,\eqref{aoiaiaskjjsss} are equivalent to the formulae (48)
for $\check{{R}}^{(1)}(\bm{w})$ and $\check{{D}}^{(1)}(\bm{w})$ quoted in ref.\cite{Bazhanov:2019xvy}.
\bigskip

To the best of our knowledge, the reflection operators for the quantum AKNS integrable structure
have not been discussed in sufficient detail in the literature.
For this reason, we'll elaborate some important points concerning them in the rest of this section.

 \subsection{$W_\infty$-algebra and  the reflection $S$-matrix for the cigar CFT\label{sec32}}

 In the case of the quantum AKNS integrable structure the r${\hat {\rm o}}$le of the
 extended conformal symmetry is played by the $W_\infty$-algebra from ref.\cite{Bakas:1991fs}.
 The latter involves an infinite set of  currents $W_j(u)$ with Lorentz spin $j=2,3,\ldots$
 satisfying the infinite system of Operator Product Expansions (OPE) of the form
 \bea\label{aiisaisa}
 W_2(u)\,W_2(0)&=&\frac{c}{2 u^4}+\frac{2}{u^2}\ W_2(0)+\frac{1}{u}\ \partial W_2(0)+O(1)
 \nonumber \\[0.2cm]
W_2(u)\,W_3(0)&=&\frac{3}{u^2}\ W_3(0)+\frac{1}{u}\ \partial W_3(0)+O(1)
 \nonumber \\[0.2cm]
 W_2(u)\,W_4(0)&=&\frac{(c+10)(17c+2)}{15 (c-2)\, u^4}
 \ W_2(0)+\frac{4}{u^2}\ W_4(0)+\frac{1}{u}\ \partial W_4(0)+O(1)
  \\[0.2cm]
W_3(u)\,W_3(0)&=&\frac{c(c+7)(2c-1) }{9(c-2)u^6}+\frac{ (c+7)(2c-1)}{3(c-2)u^4}\ 
\big(W_2(u)+W_2(0)\big)+\frac{1}{u^2}\ \Big( W_4(u)+W_4(0)\nonumber \\[0.2cm]
&+&W^2_2(u)+W^2_2(0)
-\frac{2c^2+22c-25}{30 (c-2)}\,\big (\partial^2W_2(u)+\partial^2W_2(0)\big)\Big)+O(1)
\nonumber\\[0.2cm]
&\ldots&\ .\nonumber
\eea
Note that the local field  $W_2^2$,
appearing in the last line of eq.\,\eqref{aiisaisa}
is a composite field built from the currents $W_2$ and is defined  to be the first regular term in the
 OPE  $W_2(u)W_2(0)$.
\bigskip

 When the central charge $c>2$,  $W_j=W_j(x_2+\ri x_1)$ are  holomorphic currents
 in the cigar non-linear sigma model \cite{Elitzur:1991cb,Witten:1991yr} defined on the space-time  cylinder with
 $x_1\sim x_1+2\pi$ and  $-\infty <x_2<+\infty$. This CFT admits  a dual description
  based on the Euclidean action \cite{ZAM}
 \bea\label{aoosaaso}
\tilde {\cal A}_{\rm cig}=\int_{-\infty}^\infty\rd x_2\int_0^{2\pi}\rd x_1\ \Big(\frac{1}{4\pi}\ \big[\,(\partial_a\varphi)^2+
 (\partial_a\vartheta)^2\,\big]+2\mu\ \re^{-\sqrt{k}\varphi}\ \cos\big(\sqrt{k+2}\,\vartheta\big)\,\Big)
 \eea
with the parameter $k$ related to the central charge as
\bea\label{centeq1}
c=2+\frac{6}{k}\ .
\eea
Similar to the Liouville CFT, the configuration space of the model  \eqref{aoosaaso}
 contains an asymptotic domain in which the
 interaction  term  becomes negligible and  $\varphi$, $\vartheta$ approach
 free massless fields. In particular
 \bea\label{asympeq1}
 \frac{1}{2}\,\Big(\frac{\partial}{\partial x_2}-\ri\,\frac{\partial}{\partial x_1}\Big)
 \varphi\,&\to&\, \partial\varphi_+=-\ri\sum_{m=-\infty}^{\infty}a_m\ \re^{-mu}\\
 \frac{1}{2}\,\Big(\frac{\partial}{\partial x_2}-\ri\,\frac{\partial}{\partial x_1}\Big)
 \vartheta\,&\to&\, \partial\vartheta_+=-\ri\sum_{m=-\infty}^{\infty}b_m\ \re^{-mu}\nonumber
 \eea
 and the  $\{a_m\}_{m=-\infty}^\infty$ are a set of 
creation-annihilation operators satisfying the Heisenberg algebra commutation relations
 \eqref{hssysy} and similarly for $b_m$. In terms of the asymptotic fields $\partial\varphi_+$  and
 $\partial\vartheta_+$,
 the first two $W$-currents are given by the following expressions
 \bea\label{Weq1}
&&W_2=-(\partial \varphi_+)^2-(\partial \vartheta_+)^2-\frac{1}{\sqrt{k}}\ \partial^2\varphi_+
\\[0.2cm]
&&W_3=-\ri\,
\bigg[\frac{6 k+4}{3 k}\, (\partial \vartheta_+)^3+2\,
 (\partial \varphi_+)^2\partial \vartheta_+-\sqrt{k}\, \partial^2 \varphi_+\partial\vartheta_+
+\frac{k+2}{\sqrt{k}}\,\partial\varphi_+\partial^2\vartheta_++\frac{k+2}{6 k}\, \partial^3\vartheta_+ \bigg].
 \nonumber
\eea
\bigskip

Let $P_1$ and $P_2$ be  the eigenvalues of the zero-modes $a_0$ and
$b_0$  respectively in   the Fock representation of two copies of the Heisenberg algebra,
${\cal F}_{P_1,P_2}\equiv {\cal F}^{(a)}_{P_1}\otimes {\cal F}^{(b)}_{P_2}$.
As in the Liouville CFT the Fock spaces ${\cal F}_{P_1,P_2}$  and  ${\cal F}_{-P_1,P_2}$
can be interpreted as the space of ``in'' and ``out''  (chiral) asymptotic states.
Note that, since  the potential term in \eqref{aoosaaso} becomes negligible as $\varphi\to +\infty$,
the ``in''  asymptotic space corresponds to $P_1<0$. This is opposed to  the convention used in the
 Introduction for the
Liouville CFT, as in that case it was assumed that the exponential interaction term  vanishes as the
Liouville field turns to $-\infty$.
In full analogy with the Liouville theory, one can introduce the reflection $S$-matrix for the 
cigar that intertwines the spaces of in and out asymptotic states. 
%\bea
%{\hat S}_{\rm cig}\ :\ \ {\bar {\cal F}}_{P_1,{\bar P}_2}\otimes {\cal F}_{P_1,P_2}\mapsto
%{\bar {\cal F}}_{-P_1,{\bar P}_2}\otimes {\cal F}_{-P_1,P_2}
%\eea
%(here we take into account the presence of the left chiral states 
%forming the Fock spaces ${\bar {\cal F}}_{\pm P_1,{\bar P}_2}$).
This operator admits  the factorized structure
\bea
{\hat S}_{\rm cig}=S^{(0)}_{\rm cig}\ \ {\hat{\bar s}}_{\rm cig}\otimes {\hat{s}}_{\rm cig}\ ,
\eea
where $S^{(0)}_{\rm cig}$ is a certain phase factor while 
$ {\hat{\bar s}}_{\rm cig}$ and $ {\hat{s}}_{\rm cig}$ are properly normalized operators 
acting in the chiral Fock spaces. In particular,
\bea
{\hat{s}}_{\rm cig}(P_1)\ :\ \ \ {\cal F}_{P_1,P_2}\mapsto {\cal F}_{-P_1,P_2}\ ,\ \  \
{\hat{s}}_{\rm cig}(P_1)\, |P_1,P_2\rangle=|\!-\!P_1,P_2\rangle
\eea
with $|P_1,P_2\rangle\equiv |P_1\rangle\otimes|P_2\rangle$ standing for the Fock vacuum. In the above formula
we explicitly indicate
the dependence of ${\hat{s}}_{\rm cig}$ on $P_1$, though it also depends on
$P_2$ and the parameter $k$. 
The action of the operator ${\hat{s}}_{\rm cig}$ on the excited states
is fully determined by the $W_\infty$-symmetry and
below we'll describe its construction, which is similar in spirit to that of the
Liouville reflection $S$-matrix \eqref{iaisissai} discussed in ref.\cite{Zamolodchikov:1995aa}.
\bigskip

First of all we note that the higher spin $W_j$ currents are generated through the OPE
involving the $W_j$ currents of lower spin, similar to how the current $W_4$ appears
in the singular part of $W_3(u)W_3(0)$ in eq.\,\eqref{aiisaisa}.
In fact, starting from $W_2$ and $W_3$, 
it is possible to generate all the $W_j$ by recursively computing  OPEs.
 Since the currents $W_2$ and $W_3$ can be represented 
via the Heisenberg generators using eq.\,\eqref{Weq1}, 
the Fock space ${\cal F}_{P_1,P_2}$  possesses the
 structure of the Verma module for the $W_\infty$-algebra. 
The latter is defined by means of the Fourier coefficients of $W_j(u)$:
 \bea\label{Wmodes1}
 W_{j}(u)=\sum_{n=-\infty}^\infty {\widetilde W}_j(m)\ \re^{-m u}\ .
 \eea
 Namely starting from the highest weight vector $|\, {\boldsymbol \varpi}\rangle$ satisfying the conditions
\bea
{\widetilde W}_j(m)\, |\,{\boldsymbol \varpi}\rangle=0\ ,\  \ \ \ \ \ 
{\widetilde W}_j(0)\, |\,{\boldsymbol \varpi}\rangle=\varpi_j\, |\,{\boldsymbol \varpi}\rangle\ 
\ \ \ \ \ (j=2,3;\,m=1,2,\ldots)
\eea
the Verma module ${\cal V}_{{\boldsymbol \varpi}}$ is constructed by taking all linear combinations of the basis
vectors of the form
 \bea\label{Wbasis10}
\boldsymbol{v}_I=\widetilde{W}_{2}(-i_1)\ldots\widetilde{W}_{2}(-i_m)\,
\widetilde{W}_{3}(-i_1')\ldots\widetilde{W}_{3}(-i_{m'}')|\,{\boldsymbol \varpi} \rangle\,,
\eea
where $I$ stands for the 
multi-index $I=(i_1,\ldots, i_m,i_1',\ldots,i_{m'}')$
such that $1\le i_1\le i_2\le\ldots\le i_m$ and $1\le i_1'\le i_2'\le\ldots\le i_{m'}'$.

\bigskip

In view of eq.\,\eqref{Weq1}, it is easy to see that ${\cal F}_{\pm P_1,P_2}$ is isomorphic to the
 Verma module  of the $W_\infty$-algebra  ${\cal V}_{\boldsymbol \varpi}$ with the highest weights
 ${\boldsymbol \varpi}=(\varpi_2,\varpi_3)$ related to the zero mode momenta $(P_1,P_2)$  as
\bea\label{sisoasiaso}
\varpi_2&=&P^2_1+P_2^2-\frac{1}{12}\nonumber\\[0.2cm]
\varpi_3&=&2 P_2\ \bigg(  P^2_1+\frac{3k+2}{3k}\,  P_2^2-\frac{2k+1}{12 k}\bigg)\ .
\eea
Hence the vectors $\bm{v}_I$ defined in eq.\,\eqref{Wbasis10} form a basis in 
${\cal F}_{\pm P_1,P_2}$. On the other hand, 
the Fock space contains a natural basis that is obtained by acting with the Heisenberg creation
operators on the vacuum
\be\label{Obasis10}
\bm{e}_I(P_1)=a_{-i_1}\ldots a_{-i_m}\,b_{-i_1'}\ldots b_{-i_{m'}'}|P_1,P_2\rangle\,,
\ee
where again 
$I=(i_1,\ldots, i_m,i_1',\ldots,i_{m'}')$
with $1\le i_1\le i_2\le\ldots\le i_m$ and $1\le i_1'\le i_2'\le\ldots\le i_{m'}'$.
The two bases \eqref{Wbasis10} and \eqref{Obasis10}
are, of course, linearly related:
\be\label{Umat1}
\boldsymbol{v}_J=\, \boldsymbol{e}_I(P_1)\, {\Theta^I}_J(P_1)\ .
\ee
The matrix elements of the chiral part of the cigar reflection $S$-matrix are expressed in terms
of ${\Theta^I}_J(P_1)$ as
\bea\label{aoasisaisa}
{\big[{\hat{s}}_{\rm cig}\big]^J}_I(P_1,P_2|\,Q_{\rm cig})={\Theta^J}_A(-P_1)\ {{\big[{\Theta}^{-1}\big]}^A}_I(P_1)\ .
\eea
In the l.h.s. of this equation
 we have indicated the dependence of the matrix elements of 
${\hat{s}}_{\rm cig}$  on the zero-mode momenta $(P_1,P_2)$ and the
parameter
\bea
Q_{\rm cig}\equiv -\frac{1}{\sqrt{k}}\ .
\eea
Explicit formulae for the matrix  ${\Theta^{J}}_A$ at the levels $L=1,2$ are presented in the appendix.
 
 \subsection{Local IM  and reflection operators $\check{\mathbb R}$
 in the quantum AKNS and Getmanov  integrable structures}

The CFT  \eqref{aoosaaso} admits an integrable deformation which is described by the action 
\cite{Fateev:1995ht}
\bea\label{aoosssaasok1a}
 \tilde{\cal A}_{\rm LR}=\int_{-\infty}^\infty\rd x_2\int_0^{R}\rd x_1\ \Big(\frac{1}{4\pi}\ \big[\,(\partial_a\varphi)^2+
 (\partial_a\vartheta)^2\,\big]+2\mu\ \re^{-\sqrt{k}\varphi}\ \cos\big(\sqrt{k+2}\vartheta\big)
 +\mu'\,  \re^{\frac{2\varphi}{\sqrt{k}}}\,\Big) .
 \eea
With the additional
 term the theory is not scale invariant as 
the parameter $\big(\mu^{\frac{2}{k}} \mu'\big)^{\frac{k}{2(k+2)}}$ has dimensions of
  $[mass]$. 
Thus the compactification length for the 
 space coordinate $x_1\sim x_1+R$  
 can no longer be painlessly rescaled to $2\pi$ as it was done in \eqref{aoosaaso}.
 It is believed \cite{Fateev:1995ht} that \eqref{aoosssaasok1a} provides a dual description
 for the model,
whose classical  limit is governed  by the  (Euclidean) action
\bea\label{usuasps1a}
{\cal A}_{\rm LR}=\frac{k}{4\pi}\ \int_{-\infty}^\infty\rd x_2\int_0^{R}\rd x_1\   \bigg(\, \frac{|\partial_a Z|^2}{1+|Z|^2}+\mathfrak{m}^2\, |Z|^2\,\bigg)\ \ \ \ \ \ \ \ \ \ 
(k\to+\infty)\ .
\eea
Here $Z=X+\ri Y$ is a complex field and
 the mass parameter $\mathfrak{m}\,\propto\, \big(\mu^{\frac{2}{k}} \mu'\big)^{\frac{k}{2(k+2)}}$.
The model \eqref{usuasps1a} is well known in the theory of classically integrable systems,
where it goes under the name of  the
Lund-Regge (complex sinh-Gordon I) model \cite{Pohlmeyer:1975nb,Lund:1976ze,Getmanov:1980cq}.
% It sits in a somewhat similar way to the
%AKNS classical integrable hierarchy as the usual sinh-Gordon model to the hierarchy of the KdV equation.
Furthermore, there are strong arguments in support of the integrability of the quantum theory \eqref{aoosssaasok1a}
as well.
Among other things that this implies,
it is expected that the quantum Lund-Regge model admits two sets of local IM
$\big\{{\mathbb I}_m^{\rm (LR)}\big\}_{m=1}^{+\infty},\, 
\big\{{ {\mathbb I}}_{m}^{\rm (LR)}\big\}_{m=-\infty}^{-1}$.
In the short distance limit each of these sets becomes the set of local IM 
from  the AKNS  integrable structure  associated with the  $W_\infty$-algebra.
In particular,
\bea\label{AKNSlim1a}
\lim_{R\to 0} \bigg(\frac{R}{2\pi}\bigg)^m\, {\mathbb I}_m^{\rm (LR)}
=\mathbb{I}_m^{({\rm AKNS})}\ \ \ \ \ (m=1,2,\ldots)
\eea
and  the first few representatives are expressed in terms of the $W$-currents as follows \cite{Fateev:1995ht,
Fateev:2005kx}
\bea\label{IAKNSeq1}
\mathbb{I}_1^{({\rm AKNS})}&=&\int_0^{2\pi}\frac{{\rm d}x_1}{2\pi}\ W_2 \nonumber \\[0.2cm]
\mathbb{I}_2^{({\rm AKNS})}&=&\frac{3 k}{2(3k+2)}\,\int_0^{2\pi}\frac{{\rm d}x_1}{2\pi}\ W_3 \\[0.2cm]
\mathbb{I}_3^{({\rm AKNS})}&=&
\frac{k}{(2k+1)(5k+4)}\int_0^{2\pi}\frac{{\rm d}x_1}{2\pi}\ \Big(k\,W_4+(2k+1)\,W_2^2\Big)\ .\nonumber
\eea
\bigskip

The construction of the reflection operator for the AKNS  integrable structure is based on the
observation that the theory \eqref{aoosssaasok1a} can be considered as a deformation
of the Liouville plus free massless  CFT by the term $\propto\, \mu$. This implies that the AKNS 
local IM \eqref{IAKNSeq1} should admit an alternative description in terms of the two fields
\bea\label{asdasd1a}
T(u)&=&-(\partial \varphi_+)^2+Q_{\rm L}\  \partial^2\varphi_+\ ,\ \ \ \ {\rm where}\ \  Q_{\rm L}=\frac{1+k}{\sqrt{k}}\nonumber\\
J(u)&=&\ri\, \partial\vartheta_+\ .
\eea
Indeed, it is not difficult to check that
\bea\label{AKNSint1}
{\mathbb I}_1^{({\rm AKNS})}&=&\int_0^{2\pi}\,\frac{{\rm d}x_1}{2\pi}\,\big(J^2+T\big) \nonumber \\[0.2cm]
{\mathbb I}_2^{({\rm AKNS})}&=&\int_0^{2\pi}\,\frac{{\rm d}x_1}{2\pi}\,\Big(J^3+
\frac{3 k}{3 k+2}\,J\,T\Big)  \\[0.2cm]
{\mathbb I}_3^{({\rm AKNS})}&=&\int_0^{2\pi}\frac{{\rm d}x_1}{2\pi}\ \Big(J^4
-\frac{k^2+4k+2}{5k+4}\,\big(\partial J\big)^2
+\frac{6 k}{5k+4}\ J^2\,T
+\frac{k}{5 k+4}\,T^2\Big)\ .\nonumber
\eea
Notice that in the above formulae the overall 
multiplicative normalization of the local IM is fixed such that
\be\label{AKNSnorm1}
\mathbb{I}_m^{({\rm AKNS})}=\int_0^{2\pi}\frac{{\rm d}x_1}{2\pi}\,\big(J^{m+1}+\ldots\big)\ ,
\ee
where the dots stand for terms containing  lower powers of the current $J$ and its derivatives.
\bigskip

Consider the first line in eq.\,\eqref{asdasd1a}.
The r.h.s. is identical to the expression for the
stress energy momentum tensor in the Liouville CFT with central charge
$c_{{\rm L}}=1+6\,Q^2_{{\rm L}}$, written in terms of the free Liouville asymptotic field.
%The field $T(z)$ in eq.\,\eqref{asdasd1a} coincides with the 
This suggests that the reflection operator
commuting with the local IM  \eqref{AKNSint1} is built form, together with
 $\hat{s}_{{\rm cig}}(P_1,P_2|\,Q_{\rm cig})$ given 
by eq.\,\eqref{aoasisaisa}, the operator
$\big(\hat{s}_{{\rm L}}^{(a)}\otimes\mathbf{1}_b\big)$.
The latter is to be understood as the operator 
which commutes with all the modes $\{b_m\}_{m=-\infty}^{\infty}$, and acts as 
the Liouville reflection $S$-matrix $\hat{s}_{{\rm L}}$ \eqref{iaisissai}
on vectors of the form $a_{-i_1}\,\ldots a_{-i_m}|P_1,P_2\rangle$ $(1\le i_1\le i_2\le\ldots\le i_m)$.
It turns out that
\bea\label{ajaassau}
\check{\mathbb R}^{({\rm AKNS})}
=\hat{s}_{{\rm cig}}(-P_1,P_2|\,Q_{\rm cig})\,\big(\hat{s}_{{\rm L}}^{(a)}(P_1|\,Q_{\rm L})\otimes\mathbf{1}_b\big)\,
 \ \ 
{\rm with}\ \  Q_{\rm L}=\frac{1+k}{\sqrt{k}}\, ,\ \ \ Q_{\rm cig}=-\frac{1}{\sqrt{k}}\, ,
\eea
 acts  invariantly in the Fock space ${\cal F}_{P_1,P_2}\cong {\cal V}_{\boldsymbol \varpi}$
and   commutes with the local IM \eqref{AKNSint1}:
\bea\label{kasisasis}
\big[\,\check{\mathbb R}^{({\rm AKNS})},{\mathbb I}_m^{({\rm AKNS})}\,\big]=0\ .
\eea
\vskip0.2cm

For completeness let us recall the construction of the Liouville reflection $S$-matrix $\hat{s}_{{\rm L}}(P|Q)$ \cite{Zamolodchikov:1995aa}.
%It is similar in spirit to that of the cigar reflection $S$-matrix given at the end of the previous subsection.
As was mentioned in the Introduction the Fock space  ${\cal F}_{P}$, defined as the highest weight
representation of the Heisenberg algebra \eqref{hssysy} with highest weight vector $|P\rangle$ \eqref{vac1a},
is isomorphic to the Verma module ${\cal V}_{\Delta}$ for the Virasoro algebra with $\Delta$ as in eq.\,\eqref{deltarel1}.
Hence the two different bases
\be
\bm{t}_I=L_{-i_1}\ldots L_{-i_m}|P\rangle
\ee
and
\be
\bm{e}_I(P)=a_{-i_1}\ldots a_{-i_m}|P\rangle\,,
\ee
one associated with the Virasoro algebra and the other with the Heisenberg algebra,
are linearly related.
Using
the explicit expressions for the Virasoro generators in terms of the Heisenberg ones \eqref{ososossi},
it is not difficult to read off the components of the matrix ${\Omega^I}_J(P)$
connecting the two bases,
\bea\label{hsysyas}
{\boldsymbol t}_{J}={\boldsymbol e}_I(P)\ {\Omega^I}_J(P)\ .
\eea
Then the chiral part of the Liouville reflection $S$-matrix is given by
\bea\label{jasusaus}
{\big[{\hat{s}}_{\rm L}\big]^J}_I(P|\,Q)={\Omega^J}_A(-P)\ {{\big[{\Omega}^{-1}\big]}^A}_I(P)\ .
\eea
Explicit formulae for  ${\Omega^{J}}_A$ at levels $L=1,2,3$ can be found
in the appendix.
\vskip0.2cm

The AKNS local IM \eqref{AKNSlim1a}
and the reflection operator \eqref{ajaassau} admit a heuristic interpretation 
based on the integrable QFT \eqref{aoosssaasok1a}, where $k$ is a positive coupling constant.
At the same time it is not difficult to see that for a given level $L$ the finite  matrices of the local IM,
${\mathbb I}^{\rm (AKNS)}_m$, 
as well as the matrix \eqref{ajaassau} are rational functions of $Q_{\rm cig}=-1/\sqrt{k}$  and $(P_1,P_2)$
(see, e.g., Appendix).
Hence the commutativity conditions \eqref{kasisasis}  and 
$\big[{\mathbb I}_m^{({\rm AKNS})},{\mathbb I}_{m'}^{({\rm AKNS})}\big]=0$
must remain  unchanged for any complex values of these variables. 
The first line in eq.\,\eqref{haysayas}  gives the eigenvalues of the operator \eqref{ajaassau}
provided that 
\bea\label{ksissau}
P_1=\frac{p}{\sqrt{n+2}}\  ,\ \ \ \ P_2=\frac{s}{\sqrt{n}}\ ,\ \ \ \ \sqrt{k}=-\ri\ \sqrt{n+2}\ .
\eea

%\bea\label{aoosssaaso}
% {\cal A}=\int_{-\infty}^\infty\rd x_1\int_0^{2\pi}\rd x_2\ \Bigg(\frac{1}{4\pi}\ \big[\,(\partial_a\varphi)^2+
 %(\partial_a\theta)^2\,\big]&+&2\mu\ \re^{-\sqrt{k}\varphi}\ \cos\big(\sqrt{k+2}\vartheta\big)\nonumber\\[0.2cm]
 %&+&\mu'\, \Big(\frac{R}{2\pi}\Big)^{2+\frac{4}{k}} \ \re^{\frac{2\varphi}{\sqrt{k}}}\,\Bigg)\ ,
 %\eea
 
 \bigskip
 
 Finishing this subsection let us note that the model
 \eqref{aoosaaso} admits another integrable deformation of the form
\bea\label{aoosssaasok}
 \tilde{\cal A}_{\rm G}=\int_{-\infty}^\infty\rd x_2\int_0^{R}\rd x_1\ \Big(\frac{1}{4\pi}\ \big[\,(\partial_a\varphi)^2+
 (\partial_a\vartheta)^2\,\big]+2\mu\ \re^{-\sqrt{k}\varphi}\ \cos\big(\sqrt{k+2}\vartheta\big)
 +\mu'\,  \re^{\frac{4\varphi}{\sqrt{k}}}\,\Big),
 \eea
 which gives the dual description
 for the Getmanov  (complex sinh-Gordon II) model \cite{Getmanov:1980cq}
 \bea\label{usuasps1b}
{\cal A}_{\rm G}=\frac{k}{4\pi}\ \int_{-\infty}^\infty\rd x_2\int_0^{R}\rd x_1\ 
  \bigg(\, \frac{|\partial_a Z|^2}{1+|Z|^2}+\mathfrak{m}^2\, |Z|^2\,\big(1+|Z|^2\,\big)\,\bigg)\ 
\ \ \ \ \ \ \ \ \ \ 
(k\to+\infty)\ .
\eea
Thus, as was mentioned in the Introduction, there is another integrable structure associated with the  $W_\infty$-algebra. 
The Getmanov integrable structure contains the
set of local IM $\big\{{\mathbb I}^{(\rm G)}_{2m-1}\big\}_{m=1}^{\infty}$ such that
\bea\label{Get}
\mathbb{I}_1^{({\rm G})}&=&\int_0^{2\pi}\frac{{\rm d}x_1}{2\pi}\ W_2  \\[0.4cm]
\mathbb{I}_3^{({\rm G})}&=&\frac{k^2}{4(k+2)(2k+1)(2k+3)}
\int_0^{2\pi}\frac{{\rm d}x_1}{2\pi}\ \Big(\, (k+6)\,W_4+4\,(2k+1)\,W_2^2\Big)\ .\nonumber
\eea
 The latter are related to the local IM from the massive theory \eqref{usuasps1b} similarly to the way the
AKNS local IM appear in the context of the  Lund-Regge (complex sinh-Gordon I) model  (see eq.\,\eqref{AKNSlim1a}).
The reflection operator for the Getmanov integrable structure is given by the formula analogous
to \eqref{ajaassau}
\bea\label{ajaassauss}
\check{\mathbb R}^{\rm(G)}=
\hat{s}_{{\rm cig}}(-P_1,P_2|\,Q_{\rm cig})\,\big(\hat{s}_{{\rm L}}^{(a)}(P_1|\,Q_{\rm L}')\otimes\mathbf{1}_b\big)\  \ \ \ 
{\rm with}\ \  Q_{\rm L}'=\frac{k+4}{2\sqrt{k}}\, ,\ \ \ Q_{\rm cig}=-\frac{1}{\sqrt{k}}\ .
\eea
\bigskip

\subsection{Reflection operators $\check{\mathbb C}^{(\pm)}$ and  $\check{\mathbb D}^{(\rm AKNS)}$}

Introduce a new set of holomorphic fields $(\partial \chi_+,\partial \eta_+)$ related to the basic fields
$(\partial \varphi_+,\partial \vartheta_+)$ \eqref{asympeq1} through the (complex) orthogonal transformation
\bea
\partial \chi_+ =\sqrt{\frac{k+2}{2}}\ \partial \varphi_++\ri\, \sqrt{\frac{k}{2}}\ \partial\vartheta_+\ ,\ \ \ \ 
\partial \eta_+=-\ri\ \sqrt{\frac{k}{2}}\ \partial \varphi_++\sqrt{\frac{k+2}{2}}\ \partial\vartheta_+
\eea
and  define the currents
\bea\label{aisisai}
J_1=\ri\,\partial  \chi_+\ ,\ \ \  \ J_2=-(\partial \eta_+)^2+\frac{\ri}{\sqrt{2}}\ \partial^2 \eta_+\ .
\eea
Notice that the spin-2 field $J_2$ generates the Virasoro algebra with central charge
$c=-2$. For this particular value the commuting system of local IM
$\big\{{\mathbb I}^{(c=-2)}_{2m-1}\big\}_{m=1}^\infty$ given by
\eqref{KdVint2} with $T$ substituted by  $J_2$ and $c$ set to $-2$
can  be extended  to the system $\big\{{\mathbb I}^{(c=-2)}_m\big\}_{m=1}^\infty$
with   ${\mathbb I}^{(c=-2)}_{2m}$ being the integrals over the local densities of
odd Lorentz spin built out of the field $\partial \eta_+$.
For example
\bea
{\mathbb I}^{(c=-2)}_{2}=-\ri\ \int_0^{2\pi}\frac{\rd x_1}{2\pi}\,(\partial \eta_+)^3\ .
\eea 
As was discussed in Appendix C in ref.\cite{Fateev:2005kx}, it is possible to rewrite the  local IM from the quantum AKNS integrable structure \eqref{IAKNSeq1}
in the following form
\bea\label{saooosa}
{\mathbb I}^{\rm (AKNS)}_1&=&{\mathbb I}^{(c=-2)}_{1}[\partial\eta_+]+
\int_0^{2\pi}\frac{\rd x_1}{2\pi}\ J^2_1\nonumber\\[0.2cm]
{\mathbb I}^{\rm (AKNS)}_2&=&  \frac{4 k+2}{3k+2}\ \sqrt{\frac{k+2}{2}}\  
{\mathbb I}^{(c=-2)}_{2}[\partial\eta_+]
+\frac{ \ri\, \sqrt{2k}}{3k+2} 
\int_0^{2\pi}\frac{\rd x_1}{2\pi}\ \big(k\, J_1^3+3\,(k+1)\,  J_2 J_1\,\big)\nonumber\\[0.2cm]
&\cdots&\\[0.2cm]
{\mathbb I}^{\rm (AKNS)}_m&=& c_m\ 
{\mathbb I}^{(c=-2)}_{m}[\partial\eta_+]+\int_0^{2\pi}\frac{\rd x}{2\pi}\ T_{m+1}\big(J_1,  J_2\big)\ ,
\nonumber
\eea
where $c_m$ are some $k$-dependent constants and
 $T_{m+1}$ is a certain local differential   polynomial built out of the
currents $J_1$,\,$J_2$\ \eqref{aisisai}.

\bigskip
Let ${\cal F}^{(\eta)}_{P}$ be the space of representation for the
Heisenberg operators
\bea\label{jsusssy}
\eta_m=-\ri\ \sqrt{\frac{k}{2}}\ a_m+\sqrt{\frac{k+2}{2}}\ b_m
\eea 
and 
${\hat s}_{\tt L}^{(\eta)}(P|\frac{\ri}{\sqrt{2}}\big)$ be the Liouville reflection $S$-matrix
intertwining ${\cal F}^{(\eta)}_P$ and ${\cal F}^{(\eta)}_{-P}$.
Also,
similar to \eqref{oosostr},
  introduce the  operator ${\hat C}^{(\chi)}$  of the $C$-conjugation for the Heisenberg generators
\bea
\chi_m=\sqrt{\frac{k+2}{2}}\ a_m+\ri\, \sqrt{\frac{k}{2}}\ b_m\ .
\eea
  Then we define 
\bea\label{Copeq1}
\check{\mathbb C}=
\Big[ {\hat C}^{(\chi)}\otimes{\hat s}_{\tt L}^{(\eta)}\big(-P_\eta\,\big|\,\tfrac{\ri}{\sqrt{2}}\big)\Big]\,
\Big[ {\hat s}_{\tt L}^{(a)}(P_1\,|\, Q_{\tt L})\otimes {\hat C}^{(b)}\Big]\ ,
%\ \ :\   
%{\cal F}_{P_1}\otimes {\cal F}_{P_2}\  \mapsto\  {\cal F}_{P_1}\otimes {\cal F}_{P_2}\, ,
\eea
where
\bea\label{aosaoosa}
P_\eta=-\ri\ \sqrt{\frac{k}{2}}\ P_1+\sqrt{\frac{k+2}{2}}\ P_2\ ,\ \ \ \ Q_{\rm L}=\frac{1+k}{\sqrt{k}}\ .
\eea
Notice that the second factor in the square brackets $[\,\ldots\,]$  in the r.h.s. of 
  $\eqref{Copeq1}$ acts from
${\cal F}^{(a)}_{P_1}\otimes {\cal F}^{(b)}_{P_2}$ to the space
 ${\cal F}^{(a)}_{-P_1}\otimes {\cal F}^{(b)}_{-P_2}$.
The latter is equivalent to ${\cal F}^{(\chi)}_{-P_\chi}\otimes {\cal F}^{(\eta)}_{-P_\eta}$
with $P_\eta$  given by \eqref{aosaoosa} and
$P_\chi={\sqrt{\frac{k+2}{2}}}\, P_1+\ri\sqrt{\frac{k}{2}}\, P_2$.
Since the first factor intertwines  ${\cal F}^{(\chi)}_{-P_\chi}\otimes {\cal F}^{(\eta)}_{-P_\eta}$ back to
 ${\cal F}^{(\chi)}_{P_\chi}\otimes {\cal F}^{(\eta)}_{P_\eta}\equiv
{\cal F}^{(a)}_{P_1}\otimes {\cal F}^{(b)}_{P_2} $, the operator  \eqref{Copeq1}
acts invariantly in the Fock space.
Formula  \eqref{saooosa}  suggests that
 $\check{\mathbb{C}}$ commutes with the local IM from the quantum AKNS integrable structure.

\bigskip
Up until now $k$ was assumed to be a positive real number and $\sqrt{k}$, $\sqrt{k+2}$
were understood to be the arithmetic square roots.
The analytic continuation of the operator \eqref{Copeq1} to the domain $k<-2$
using eq.\,\eqref{ksissau} with $n>0$ requires one to specify the 
branch of $\sqrt{k+2}$.
Let us set
\be\label{kplus2eq1}
\sqrt{k+2}=-\ri\sqrt{n}\qquad \qquad (\sqrt{n}>0)\ .
\ee
This results in the operator
\bea\label{Cpluseq1}
\check{\mathbb C}^{(+)}=
\Big[{\hat C}^{(\chi)}\otimes {\hat s}_{\tt L}^{(\eta)}
\big(\tfrac{p+ \ri \, s}{\sqrt{2}}\,\big|\,\tfrac{\ri}{\sqrt{2}}\big)\Big]\ 
\Big[{\hat s}_{\tt L}^{(a)}\big(\tfrac{p}{\sqrt{n+2}}\,\big|\, 
\tfrac{n+1}{\ri\sqrt{n+2}}\big)\otimes{\hat C}^{(b)} \Big]\ .
\eea
If one were to choose the other branch of the square root in eq.\,\eqref{kplus2eq1}
one would obtain the operator $\check{\mathbb{C}}^{(-)}$. It is 
easy to see that 
\be\label{Cminuseq1}
\check{\mathbb{C}}^{(-)}=
\big(\mathbf{1}_a\otimes {\hat C}^{(b)} \big)\,{\mathbb C}^{(+)}\big(\mathbf{1}_a\otimes {\hat C}^{(b)} \big)\ .
\ee
The eigenvalues of $\check{\mathbb{C}}^{(\pm)}$ are given by 
$C_p^{(\pm,L)}/C_p^{(\pm,0)}\big|_{n_a=n}$ with $C_p^{(\pm,L)}$ defined by 
eqs.\,\eqref{hsaysayt},\,\eqref{aoisisaiaskjj}.
The operators $\check{\mathbb{R}}^{({\rm AKNS})}$ and $\check{\mathbb{D}}^{({\rm AKNS})}$
with eigenvalues 
 \eqref{RDeig1a}, are simply expressed in terms of $\check{\mathbb{C}}^{(\pm)}$:
\be\label{RDCPeq1}
\check{\mathbb{R}}^{({\rm AKNS})}=\check{\mathbb{C}}^{(+)}\,\check{\mathbb{C}}^{(-)}\,,\qquad
\check{\mathbb{D}}^{({\rm AKNS})}=\check{\mathbb{C}}^{(+)}\,\big[\check{\mathbb{C}}^{(-)}\big]^{-1}\ .
\ee
Combining the above with eqs.\,\eqref{Cpluseq1},\,\eqref{Cminuseq1} yields
\be\label{iasususau}
\check{\mathbb{D}}^{({\rm AKNS})}=
\Big[{\hat C}^{(\chi)}\otimes {\hat s}_{\tt L}^{(\eta)}
\big(\tfrac{p+ \ri \, s}{\sqrt{2}}\,\big|\,\tfrac{\ri}{\sqrt{2}}\big)\Big]
\big(\mathbf{1}_a\otimes {\hat C}^{(b)} \big)
\Big[ {\hat C}^{(\chi)}
\otimes{\hat s}_{\tt L}^{(\eta)}
\big(-\tfrac{p-\ri \, s}{\sqrt{2}}\,\big|\,\tfrac{\ri}{\sqrt{2}}\big)\Big]
\big(\mathbf{1}_a\otimes {\hat C}^{(b)} \big)\ .
\ee
%\be\label{iasususau}
%\check{\mathbb{D}}^{({\rm AKNS})}=
%\Big[\mathbf{1}_{\chi}\otimes\, \hat{C}^{(\eta)} {\hat s}_{\tt L}^{(\eta)}
%\big(-\tfrac{p+ \ri \, s}{\sqrt{2}}\,\big|-\tfrac{\ri}{\sqrt{2}}\big)\Big]
%\big( {\hat C}^{(a)}\otimes\mathbf{1}_b \big)
%\Big[\mathbf{1}_{\chi}
%\otimes\, \hat{C}^{(\eta)}{\hat s}_{\tt L}^{(\eta)}
%\big(\tfrac{p-\ri \, s}{\sqrt{2}}\,\big|-\tfrac{\ri}{\sqrt{2}}\big)\Big]
%\big({\hat C}^{(a)}\otimes\mathbf{1}_b  \big)\ .
%\ee
At the same time, due to the relation \eqref{ajaassau},
it is possible to use \eqref{Cpluseq1}-\eqref{RDCPeq1}
to express the cigar reflection $S$-matrix in terms of the Liouville one.
A straightforward calculation leads to the remarkable formula
\bea\label{ScigLiouveq1}
\hat{s}_{{\rm cig}}(P_1,P_2\,|\,Q\big)&=&
\Big[\, {\hat C}^{(\chi)}\otimes {\hat s}_{\tt L}^{(\eta)}
\Big(\tfrac{\ri }{\sqrt{2}\,Q}\,P_1-
\tfrac{\sqrt{1+2Q^2}}{\sqrt{2} \,Q}\,P_2\,\Big|\,\tfrac{\ri}{\sqrt{2}}\Big) \,\Big]\
\Big[\,{\hat s}_{\tt L}^{(a)}\big(-P_1\,\big|
-Q-Q^{-1}\big)\otimes \bm{1}_b\, \Big]\nonumber\\[0.2cm]
&\times&
\Big[\,{\hat C}^{(\chi)}\otimes {\hat s}_{\tt L}^{(\eta)}
\Big(\tfrac{\ri }{\sqrt{2}\,Q}\,P_1+
\tfrac{\sqrt{1+2Q^2}}{\sqrt{2} \,Q}\,P_2\, \Big|\,\tfrac{\ri}{\sqrt{2}}\Big)\,\Big]\ .
\eea
Here $P_1$, $P_2$ and $Q$ can be taken to be arbitrary complex numbers, while the 
set of Heisenberg generators $\{\eta_m,\chi_m\}_{m\ne 0}$ are related to the set
$\{a_m,b_m\}_{m\ne 0}$ as
\bea
\eta_m&=&\tfrac{\ri}{\sqrt{2}\,Q}\ a_m+
\tfrac{\sqrt{1+2Q^2}}{\sqrt{2} \,Q}\ b_m  \\[0.2cm]
\chi_m&=&\tfrac{\sqrt{1+2Q^2}}{\sqrt{2} \,Q}\ a_m-\tfrac{\ri}{\sqrt{2}\,Q}\ b_m\ . \nonumber
\eea

 \section{\label{sfour}Spectrum of the reflection operator in the  paperclip  integrable structure}

 As was mentioned in the Introduction there are at least three different integrable structures
associated with the  $W_\infty$-algebra. Two of them are the AKNS and Getmanov integrable
structures that were discussed in the previous section.
The third one is related to the so-called sausage model \cite{Fateev:1992tk},
whose classical action reads as follows
  \bea\label{usuasps}
{\cal A}_{\rm saus}=\frac{k}{4\pi}\    \int_{-\infty}^\infty\rd x_2\int_0^{R}\rd x_1\
\frac{(1-\lambda)\ |\partial_a Z|^2}{(1+\lambda |Z|^2)(1+|Z|^2) }\ 
\ \ \ \ \ \ \ \ \ \ 
(k\to+\infty)\ ,
\eea
where $0<\lambda<1$ is some constant.\footnote{%
The mass scale in the quantum theory \eqref{usuasps} appears through the 
mechanism of dimensional transmutation of the bare coupling 
$\lambda$.}
The dual description of the sausage model 
was originally proposed by Al. Zamolodchikov \cite{ZAM} and can also be considered as
a massive integrable deformation of \eqref{aoosaaso}:
 \bea\label{aoosssaasokss}
 \tilde{\cal A}_{\rm saus}=\int_{-\infty}^\infty\rd x_2\int_0^{R}\rd x_1\ \Big(\frac{1}{4\pi}\,\big[\,(\partial_a\varphi)^2+
 (\partial_a\theta)^2\,\big]+4\mu\, \cosh\big(\sqrt{k}\,\varphi\big) \cos\big(\sqrt{k+2}\,\vartheta\big)
\Big)\, .
 \eea
In the short distance limit the set of local IM for the massive theory become
$\big\{\mathbb{I}_{2m-1}^{({\rm pc})}\big\}_{m=1}^\infty$, whose first few members
are given in terms of the $W_{\infty}$-currents as  \cite{Fateev:1995ht,Lukyanov:2003nj}
 \bea\label{PCIM1a}
\mathbb{I}_1^{({\rm pc})}&=&\int_0^{2\pi}\frac{{\rm d}x_1}{2\pi}\ W_2  \\[0.2cm]
\mathbb{I}_3^{({\rm pc})}&=&\frac{k}{(k+2)(2k+1)(3k+2)}
\int_0^{2\pi}\frac{{\rm d}x_1}{2\pi}\ \Big(\, k\,W_4+(2k+1)(3k+4)\,W_2^2\Big)\ .\nonumber
\eea
These local IM play an important r$\hat{\rm{o}}$le in the description of the 
so-called paperclip  boundary state  \cite{Lukyanov:2003nj} and we use the 
superscript ``(pc)'' for their notation.
The associated reflection operator, which commutes with the
paperclip IM, can be expressed in the form 
\bea\label{ajaassauss}
\big(\check{\mathbb R}^{({\rm pc})}\big)^2=
\big[\hat{s}_{{\rm cig}}(-P_1,P_2|\,-Q_{\rm cig})
\,\hat{s}_{{\rm cig}}(P_1,P_2|\,Q_{\rm cig})\big]^{-1}\ \ \ \ \ 
{\rm with}\ \  \ \ \ Q_{\rm cig}=-\frac{1}{\sqrt{k}}\ .
\eea
The  reason why we define the r.h.s. to be the square of the reflection operator is the following.
It is not difficult to check that the simultaneous change of sign of $P$ and $Q_{\rm cig}$ corresponds
to the $C$-conjugation of $\hat{s}_{{\rm cig}}$:
\be
\hat{s}_{{\rm cig}}(-P_1,P_2|\,-Q_{\rm cig})=(\hat{{ C}}^{(a)}\otimes\bm{1}_b)\,\hat{s}_{{\rm cig}}
(P_1,P_2\,|\,Q_{\rm cig})\,(\hat{{ C}}^{(a)}\otimes\bm{1}_b)
\ee
(recall that  $(\hat{{ C}}^{(a)}\otimes\bm{1}_b)$ acts as in \eqref{oosostr} on the modes $a_m$ and 
 as the identity operator on the $b$-modes).
From this it immediately follows that
$\check{\mathbb R}^{({\rm pc})}$ can be written in the form similar to \eqref{sosoisis}
\be\label{RPC1a}
\check{\mathbb R}^{({\rm pc})}=
\big[(\hat{{ C}}^{(a)}\otimes\bm{1}_b)\,\hat{s}_{{\rm cig}}(P_1,P_2|\,Q_{\rm cig})\big]^{-1}\, .
\ee
The spectrum of the reflection operator \eqref{RPC1a} is obtained from the general result  
quoted in sec.\,\ref{stwo}
by the following specialization of the parameters (for details, see ref.\!\cite{Bazhanov:2017nzh})
\bea
p_1=-\ri\, \sqrt{k}\, P_1,\ \ \ \ p_2=\sqrt{k+2}\, P_2\ ,\ \ p_3=0\ ,\ \ \ \ a_1=-k\ ,\ \ \ \ a_2=k+2\ .
\eea
Notice that in this case the   system of algebraic  equations
\eqref{spossaopa},\,\eqref{sooad},\,\eqref{aiisiss} for
${\boldsymbol x}=(x_1,\ldots, x_L)$  is expected to possess
${\tt par}_2(L)$ different (up to permutations) solutions, where ${\tt par}_2(L)$ is
the number of bipartitions of $L$.
Each solution ${\boldsymbol x}$ corresponds to a  certain eigenvector 
$|{\boldsymbol x}\rangle_{\boldsymbol p}
\in{\cal F}_{P_1,P_2}$ of the reflection operator \eqref{RPC1a}, and the corresponding eigenvalue
$\check{R}^{(\rm pc)}({\boldsymbol x})$ is such that
\bea
\big(\check{R}^{(\rm pc)}({\boldsymbol x})\big)^2
&=&G_L\big({\boldsymbol x}\,|-\ri \sqrt{k}\ P_1,+\sqrt{k+2}\, P_2,0\!\parallel\!-k,k+2\big)\, \nonumber\\[-0.2cm]
&& \\[-0.2cm]
&\times & G_L\big({\boldsymbol x}\,|-\ri\sqrt{k}\ P_1,-\sqrt{k+2}\, P_2,0\!\parallel\!-k,k+2\big)\ .\nonumber
\eea
Here the function $G_L({\boldsymbol x}\,|\,p_1,p_2,p_3\!\parallel\!a_1,a_2)$ is defined in
eqs.\,\eqref{isisisai},\,\eqref{aoisisaias}.

 \section{\label{sfive}Spectrum of the reflection operator in the quantum KdV integrable structure}

 The local IM in the quantum KdV integrable structure \eqref{KdVint2} appear 
  in  the short distance limit of the quantum sinh-Gordon model
 \bea\label{sGaction1}
  {\cal A}_{\rm shG}=\int_{-\infty}^\infty\rd x_2\int_0^{R}\rd x_1\ \Big(\frac{1}{4\pi}\, (\partial_a\varphi)^2
 +2\mu\, \cosh\big(2b\varphi\big) \,\Big)\ .
 \eea
 Similar to eq.\,\eqref{AKNSlim1a} one has
 \bea\label{Slim1a}
\lim_{R\to 0} \bigg(\frac{R}{2\pi}\bigg)^{2m-1}\, {\mathbb I}_{2m-1}^{\rm (shG)}
=\mathbb{I}_{2m-1}^{({\rm KdV})}\ \ \ \ \ \ \ \ \  \ \ \  (m=1,2,\ldots)\ .
\eea
 Here the relation between the coupling constant $b$ entering \eqref{sGaction1}
 and $Q$, which parameterizes $c$ in eq.\,\eqref{KdVint2}
as $c=1+6\,Q^2$, is given by
 \bea
 Q=b^{-1}+b\ .
 \eea
 The local  IM  $\big\{{\mathbb I}^{({\rm KdV})}_{2m-1}\big\}^\infty_{m=1}$ act invariantly in the
 level subspace of the Fock space ${\cal F}_P$ and commute
 with the operator
 $
 \hat{s}_{{\rm L}}(-P|\,-Q)
\,\hat{s}_{{\rm L}}(P|\,Q)
 $.
 Taking into account  the relation
 \bea
 \hat{s}_{{\rm L}}(-P|\,-Q)={\hat C}\,\hat{s}_{{\rm L}}(P|\,Q)\,{\hat C}\,,
 \eea
 where $\hat{C}$ denotes the $C$-conjugation \eqref{oosostr}, one concludes that the reflection 
 operator in eq.\eqref{sosoisis} commutes with all members of the set
 $\big\{{\mathbb I}^{({\rm KdV})}_{2m-1}\big\}^\infty_{m=1}$. Its eigenvalues 
 $R^{({\rm KdV})}({\bm v})$ \eqref{hassayyas} follow from the results of sec.\,\ref{sthreeone}.
They can be obtained through a certain reduction of
 eqs.\,\eqref{haysayas}-\eqref{sksksk1}
that will be described below.

 \bigskip
Consider the algebraic system \eqref{sksksk1} in the case with even
  $L=2N$ and $s=0$. It  admits  solutions  such that 
\bea\label{aiisai}
w_{2N+1-a}=-w_a\qquad {\rm with} \qquad  a=1,\ldots, N\ .
\eea
Using  the set $\{v_a\}_{a=1}^N$ defined by the formula
\bea\label{jsausausua}
w_a^2=-\frac{(n+2)^2}{2n}\ v_a\ \ \ \ \ \ \ \ \ \ \ \ \ \  \ (a=1,\ldots, N)\ ,
\eea
 equations   \eqref{sksksk1}  can be rewritten in the form \eqref{jsaysssysa},
 provided that the parameters are identified as follows
\bea
\alpha=-\frac{n}{n+2}\ ,\ \ \ \ \Delta=\frac{4\,p^2-n^2}{8(n+2)}\ .
\eea
In connection with this, let us note that the above reduction brings the ODE 
\bea\label{ODE2a}
\bigg[-\frac{\rd^2}{\rd w^2}+\frac{p^2-\frac{1}{4}}{w^2}+\frac{2\ri s}{w}+1+\sum_{a=1}^L\bigg(
\frac{2}{(w-w_a)^2}+\frac{n}{w(w-w_a)}\,\bigg)+\lambda^{-2-n}\ w^n\,\bigg]\ \Psi=0\ ,
\eea
which plays the central r${\hat{\rm o}}$le in the quantum AKNS integrable structure (see  \cite{Bazhanov:2019xvy} for details),
to the form
\bea\label{aisausau}
\Bigg[\,-\frac{\rd ^2}{{\rd w}^2}+\frac{p^2-\frac{1}{4}}{w^2}+1+
\sum_{a=1}^N\bigg(\frac{4 (w^2+w_a^2)}{(w^2-w^2_a)^2}
+\frac{2n}{w^2-w^2_a}\bigg)+\lambda^{-2-n}\ w^n\,
\Bigg]\, \Psi=0\ .
\eea
The  latter  is equivalent to the  Schr${\ddot{\rm o}}$dinger   equation with Monster potentials
 associated with the quantum KdV integrable structure \cite{Bazhanov:2003ni}.
 Indeed,  the change of variables 
\bea
\Psi(w)
=
%y^{-\frac{n}{2(n+2)}}\ {\tilde \Psi}(y)
y^{\frac{\alpha}{2}}\ {\tilde \Psi}(y)
\ ,\ \ \ \ \  \ w=
%\frac{n+2}{2}\ y^{\frac{2}{n+2}}=
\frac{y^{\alpha+1}}{\alpha+1}
\eea
transforms the ODE  \eqref{aisausau} to 
\bea\label{aisiasaiasi}
\bigg(\,-\frac{\rd ^2}{{\rd y}^2}+V_{\rm Monst}(y)+E\, \bigg)\, {\tilde \Psi}(y)=0\ ,
\eea
where
\bea
V_{\rm Monst}(y)=
\frac{{ \ell}({ \ell}+1)}{y^2}+y^{2\alpha}-
2\ \frac{\rd^2}{\rd y^2}\sum_{a=1}^{N}\log\big(y^{2\alpha+2}-\tfrac{\alpha+1}{\alpha}\,  v_a\big)\ .
\eea
Here the parameters of the Schr${\ddot{\rm o}}$dinger operator are related to those of the original 
ODE \eqref{aisausau} as
\bea\label{jashsaas}
E=\frac{4}{(n+2)^2}\ \bigg(\frac{n+2}{2\lambda}\bigg)^{n+2}\ ,\ \ \ \ 
 \ \ \ { \ell }=\frac{2p}{n+2}-\frac{1}{2}\ ,\ \ \ \ \ \ \alpha=-\frac{n}{n+2}\ ,
\eea
while the set $\{v_a\}_{a=1}^N$ satisfies \eqref{jsaysssysa} with $\Delta=\frac{(2\ell+1)^2-4\alpha^2}{16(\alpha+1)}$.

\bigskip
The specialization of \eqref{aoiaiaskjjsss} to the case \eqref{aiisai},\eqref{jsausausua}
gives
\bea\label{aoiaiasss}
V^{(\pm)}_a(D)
&=&  (D-1)^2-
 \bigg( 2p+2+n\mp 2\ri\, (n+2)\  \sqrt{\frac{ v_a}{2n}}+2+
\sum_{b\not= a}^N\frac{8 v_a}{v_a-v_b}\bigg)\ (D-1)\nonumber\\[0.2cm]
&+&\tfrac{1}{2}\, n^2+\big(p+\tfrac{3}{2}\big)\ n\mp \ri\,(n+1+2p)\, (n+2)\  \sqrt{\frac{ v_a}{2n}}+2 p+1
\\[0.2cm]
&+&\bigg(1+\sum_{b\not=a}^N\frac{4 v_a}{v_a-v_b}\,\bigg)^2
+\, \bigg(\,
4p+2\mp 4\,\ri\, (n+2)\  \sqrt{\frac{ v_a}{2n}}+n \,\bigg)\,
\bigg(\frac{1}{2}+\sum_{b\not=a}^N\frac{2v_a}{v_a-v_b}\bigg)\ .\nonumber
\eea
Now we define
\bea
V_{a}(D)=
\begin{cases}
V^{(+)}_a(D)\ \ \ \ &{\rm for}\ \ \ a=1,\ldots, N\\[0.2cm]
V^{(-)}_{2N+1-a}(D)\ \ \ \ &{\rm for}\ \ \ a=N+1,\ldots, 2N
\end{cases}
\eea
and
\bea
w_{a}=
\begin{cases}
+\ri\, (n+2)\  \sqrt{\frac{ v_a}{2n}}\ \ \ \ &{\rm for}\ \ \ a=1,\ldots, N\\[0.2cm]
-\ri\, (n+2)\  \sqrt{\frac{ v_{2N+1-a}}{2n}}\ \ \ \ &{\rm for}\ \ \ a=N+1,\ldots, 2N
\end{cases}\ .
\eea
Then the eigenvalues $\check{R}^{({\rm KdV})}({\boldsymbol v})$  
of the normalized reflection operator   in the quantum  
KdV integrable structure, 
\bea
\check{\mathbb R}^{({\rm KdV})}=\big[{\hat  C}{\hat s}_{\rm L}(P|\,Q)\big]^{-1}\ ,
\eea
in the $N^{\rm th}$-level subspace ${\cal F}_P^{(N)}$ of the Fock space
with
\bea
P=-\frac{p}{\sqrt{2 (n+2)}}\ ,\quad \quad Q=-\frac{\ri\, n}{\sqrt{2(n+2)}}
\eea
are given by 
\bea\label{hasayas}
\check{R}^{({\rm KdV})}({\boldsymbol v})=\frac{ \det\big(w_a^{b-1}\,{V}_a(b)\big)}
{\prod_{a=1}^{2N} w_a\ \prod_{b>a}(w_b-w_a)\ \prod_{a=1}^{2N}\big(2p+2a-1\big)}\ .
\eea

In ref.\cite{Kotousov:2019ygw} the eigenvalues $\check{R}^{({\rm KdV})}$
were quoted for the level $N=1$ and $N=2$ in eqs.(7.9) and (7.10) respectively.
The expressions are equivalent to formula \eqref{hasayas}
provided that the parameters in that work are identified with those in the current paper as
 $\beta=\sqrt{\frac{n+2}{2}}$, $\rho=-\frac{n}{\sqrt{2(n+2)}}$ 
and the notation $p$ from \cite{Kotousov:2019ygw} coincides with $P$.
%rt{\frac{n+2}{2}}$, $\rho=-\frac{n}{\sqrt{2(n+2)}}$ and the notation $p$ from that work coincides with $P$
%in the current paper. 
Also in \cite{Kotousov:2019ygw} 
a simple formula was presented for 
the product of the eigenvalues for a given level $N$, i.e., 
${\rm det}_N\big(\check{\mathbb R}^{(\rm KdV)}\big)$.
It reads as
\bea\label{deteq1a}
{\rm det}_N \big(\check{\mathbb R}^{(\rm KdV)}\big)
=
%\prod_{\boldsymbol w^{(N)}}\big(\check{\mathbb R}^{(\rm KdV)}\big)=
 \prod_{1\leq j,m\leq N\atop jm\leq N}\bigg[\frac{2P+m\beta^{-1}-j\beta }
 {2P-m\beta^{-1}+j\beta }\bigg]^{{\tt par}_1(N-mj)}\,\ \ \ \ \ \ \ \ 
 \bigg(\beta=\sqrt{\frac{n+2}{2}}\ \bigg)\ ,
\eea
where ${\tt par}_1(N)$ is the number of integer partitions of $N$.

\bigskip
Significant simplifications occur for
the quantum  
KdV integrable structure when the central charge $c=-2$.
 In this case all the eigenvalues \eqref{hasayas} take the form
 \bea\label{iissaisaisa}
 \check{R}^{(c=-2)}({\boldsymbol v})=\prod_{j=1}^J\frac{\sqrt{2}\, P+n^{(-)}_j-\half}{\sqrt{2}\, P-n^{(+)}_j+\half}\ .
 \eea
Here $\big\{n_j^{(\pm)}\big\}$ are two sets of integers satisfying the conditions
 $1\le n_1^{(\pm)}<n_2^{(\pm)}<\ldots <n_J^{(\pm)}$ and 
\be
N=\sum\limits_{j=1}^J\,\big(n^{(+)}_j+n^{(-)}_j-1\big)\ .
\ee
In fact, the sets $\big\{n_j^{(\pm)}\big\}$ can be used to classify  the  states $|{\boldsymbol v}\rangle_P$ 
for any $c\leq 1$. The integers which appear in the
exact Bohr-Sommerfeld quantization condition
for the  Schr$\ddot{\rm o}$dinger equation with the Monster potentials \eqref{aisiasaiasi}  
are expressed through these numbers
(for details, see Appendix A in ref.\!\cite{Bazhanov:2003ni}).

Equation  \eqref{deteq1a} or/and \eqref{iissaisaisa}  combined with the formula \eqref{iasususau} explains the simple
form of the determinant  of the operator $\check{\mathbb{D}}^{({\rm AKNS})}$ in the
level subspaces $L=1,2$ which appear in eqs.(80),(81) from ref.\cite{Bazhanov:2019xvy}.
It is possible to show that for general $L$
\bea
\!\!
{\rm det}_L \big(\check{\mathbb D}^{(\rm AKNS)}\big)=
\prod_{N=1}^L\!
 \prod_{1\leq j,m\atop jm\leq N}\!
\bigg[\frac{(2p-2\ri s+2m-j)\,(2p+2\ri s-2m+j) }
{(2p+2\ri s+2m-j)\,(2p-2\ri s-2m+j) }\bigg]^{{\tt par}_1(N-mj)\,{\tt par}_1(L-N)}
\eea

\bigskip
It would be remiss not to mention the other integrable structure
associated with the Virasoro algebra.
The first representatives from the corresponding set of local IM,  $\big\{{\mathbb I}^{({\rm BD})}_{6m-5},
{\mathbb I}^{({\rm BD})}_{6m-1}\big\}_{m=1}^\infty$,
are given by
\bea
{\mathbb I}_1^{({\rm BD})}&=&\int_0^{2\pi}\frac{{\rm d}x_1}{2\pi} \ T=
\int_0^{2\pi}\frac{{\rm d}x}{2\pi} \ \tilde{T}\\[0.2cm]
{\mathbb I}_5^{({\rm BD})}&=&\int_0^{2\pi}\frac{{\rm d}x_1}{2\pi} \ \Big(T^3+\frac{8-\tilde{c}}{8}\ \big(\partial T\big)^2\Big)
=\int_0^{2\pi}\frac{{\rm d}x_1}{2\pi} \ \Big(\tilde{T}^3+
\frac{8-c}{8}\ \big(\partial \tilde{T}\big)^2\Big)\ ,\nonumber 
\eea
where
\be
T= -(\partial\varphi_+)^2+ Q\,\partial^2\varphi_+\ , \ \ \ \ \ 
\tilde{T}= -(\partial\varphi_+)^2+ \tilde{Q}\,\partial^2\varphi_+\ .
\ee
 Each of the holomorphic fields $T$  and ${\tilde T}$ generates the Virasoro
algebra with different central charges $c=1+6 Q^2$ and $\tilde{c}=1+6 \tilde{Q}^2$ respectively. The latter are not 
independent, but satisfy the  quadratic relation
\be
4\,(\tilde{c}^2+c^2)-17\,\tilde{c}c+117\,(\tilde{c}+c)+504=0
\ee
so that parameterizing $Q$ as $Q=b+b^{-1}$, the tilde counterpart is given by
$\tilde{Q}=-2b+(-2b)^{-1}$.
The above integrable structure is related to the Bullough-Dodd model described by the action
\bea
  {\cal A}_{\rm BD}=\int_{-\infty}^\infty\rd x_2\int_0^{R}\rd x_1\ \Big(\,\frac{1}{4\pi}\, 
  (\partial_a\varphi)^2+
 \mu\, \re^{2b\varphi}+\tilde{\mu}\, \re^{-4b\varphi} \,\Big)\ .
 \eea
One can introduce the reflection operator for the Bullough-Dodd integrable structure,
commuting with the local IM,
as
\be
\check{\mathbb R}^{({\rm BD)}}=\hat{s}_{{\rm L}}(-P\,|\,{\tilde Q})\,\hat{s}_{{\rm L}}({P}\,|\, { Q})\ .
\ee

%with
%\bea\label{isisisai}
%\frac{R_p^{(0,{\rm KdV})}}{=2^{-p-\frac{1}{2}}\ \frac{\Gamma(1+2p)}{\Gamma(\frac{1}{2}+p)}\ .
%\eea

 \section{\label{ssix}
Reflection operators and Hermitian structures
}
Up till now we have been focused on the spectral problem
for the different commuting families of operators and, as such, 
there has not been any mention of the
Hermitian structures consistent with these integrable structures.
By consistent, among other things, we take to mean that
with respect to the
formal Hermitian conjugation in the algebra of extended conformal
symmetry, the (properly normalized) local IM
are Hermitian operators. 
For the integrable structures considered in this paper
there are several natural Hermitian structures, which are related
to one another via the reflection operators.
Let us first illustrate this for the simplest case of
the KdV integrable structure, following ref.\cite{Kotousov:2019ygw}.
\bigskip

The Virasoro algebra commutation relations \eqref{sosososoy} 
admit the natural Hermitian conjugation given by
\be\label{asdaskk1}
L_{m}^\star=L_{-m}\ .
\ee
It is not difficult to see that $\big[T(u)\big]^\star=T(-u^*)$ and hence
the KdV IM \eqref{KdVint2} are Hermitian operators with
respect to the conjugation \eqref{asdaskk1}.
On the other hand, in view of eq.\,\eqref{ososossi} that
expresses the Virasoro generators in terms of the Heisenberg ones,
we can consider another conjugation
\be\label{asdaskk2}
a_{m}^\dag=a_{-m}
\ee
that is consistent with the commutation relations \eqref{hssysy}.
It is easy to see from eq.\,\eqref{ososossi} that for real $Q$, i.e.,
$c\ge 1$, the dagger conjugation of $L_m$ is identical to \eqref{asdaskk1}. However as $c<1$ 
this conjugation acts highly non-trivially on the Virasoro generators.
Nevertheless, it is possible to show that 
the KdV local IM are Hermitian w.r.t. the dagger conjugation as well (see, e.g., \cite{Kotousov:2019ygw})
\be
\Big[\mathbb{I}_{2m-1}^{({\rm KdV})}\Big]^\dagger=\Big[\mathbb{I}_{2m-1}^{({\rm KdV})}\Big]^\star=
\mathbb{I}_{2m-1}^{({\rm KdV})}\ .
\ee
\smallskip

Each of the conjugations \eqref{asdaskk1}
and \eqref{asdaskk2} can be used to introduce a Hermitian form in the space
${\cal F}_P$ for real $P$. Namely, for any vectors $\bm{\psi}_{1,2}\in{\cal F}_P$,
the forms ${\bf V}(\bm{\psi}_2,\bm{\psi}_1)$ and ${\bf H}(\bm{\psi}_2,\bm{\psi}_1)$
are defined uniquely by
\be
{\bf V}(\bm{\psi}_2,L_n\bm{\psi}_1)={\bf V}(L_{-n}\bm{\psi}_2,\bm{\psi}_1)\,,\qquad
{\bf H}(\bm{\psi}_2,a_n\bm{\psi}_1)={\bf H}(a_{-n}\bm{\psi}_2,\bm{\psi}_1)
\ee
along with the normalization condition
\be\label{norm1a}
{\bf V}(\bm{\psi},\bm{\psi})={\bf H}(\bm{\psi},\bm{\psi})=1\qquad {\rm for} \qquad \bm{\psi}=|P\rangle\ .
\ee
It turns out that  ${\bf V}$ and ${\bf H}$ coincide
when the central charge $c\ge1$.
However for $c<1$, as was pointed out in the work \cite{Kotousov:2019ygw},
they are related through the reflection operator as
\be\label{popopsa1}
{\bf H}\big(\bm{\psi}_2,\bm{\psi}_1\big)={\bf V}\big(\bm{\psi}_2,\check{\mathbb{R}}^{({\rm KdV})}\bm{\psi}_1\big)\ .
\ee
\smallskip

For the $W_\infty$-algebra, whose commutation relations are encoded by the
infinite set of OPEs \eqref{aiisaisa}, the natural Hermitian conjugation,
similar to \eqref{asdaskk1}, reads as
\be\label{Wconjeq2}
\big[\widetilde{W}_j(m)\big]^\ddag=\widetilde{W}_j(-m)\ .
\ee
Since the $W_\infty$-algebra is bosonized by means of two copies of the
Heisenberg algebra, the analogue of the dagger conjugation \eqref{asdaskk2}
is now
\be\label{aslql1}
a_{m}^\dagger=a_{-m}\,,\qquad b_m^\dagger=b_{-m}\ .
\ee
For the case with the central charge $c\ge 2$, i.e., the parameter $k$ in \eqref{centeq1} 
is real positive, the
two conjugations coincide and the AKNS and paperclip local IMs are Hermitian w.r.t.
both of them. The situation is more complicated when $c<2$.
It is easy to see from eqs.\,\eqref{Weq1},\,\eqref{IAKNSeq1} that when 
$k$ is negative or more generally a complex number, 
$\mathbb{I}_2^{({\rm AKNS})}$ cannot possibly be Hermitian under the conjugation \eqref{aslql1}.
At the same time, as it follows from the discussion in ref.\cite{Lukyanov:2003nj},
it turns out that the system of paperclip local IM \eqref{PCIM1a} are still Hermitian 
under the dagger conjugation when $k$ is a negative real number.
For $c<2$, i.e., for real negative $k$ 
we found that:
\be\label{form1as}
{\bf H}\big(\bm{\psi}_2,\bm{\psi}_1\big)={\bf W}\big(\bm{\psi}_2,\check{\mathbb{R}}^{({\rm pc})}\bm{\psi}_1\big)\ .
\ee
Here the Hermitian form ${\bf H}$ is defined through the relation
\be
{\bf H}(\bm{\psi}_2,a_{m}\bm{\psi}_1)={\bf H}(a_{-m}\bm{\psi}_2,\bm{\psi}_1)\,,\qquad
{\bf H}(\bm{\psi}_2,b_{m}\bm{\psi}_1)={\bf H}(b_{-m}\bm{\psi}_2,\bm{\psi}_1)\,,
\ee
while for ${\bf W}$ the corresponding equation is
\be
{\bf W}\big(\bm{\psi}_2,\widetilde{W}_j(m)\bm{\psi}_1\big)={\bf W}\big(\widetilde{W}_j(-m)\bm{\psi}_2,\bm{\psi}_1\big)
\qquad
(j=2,3,\ldots)\ .
\ee
In the above formulae the vectors $\bm{\psi}_{1,2}\in{\cal F}_{P_1,P_2}$ with real $P_1$, $P_2$
and the Hermitian forms are assumed to be normalized as
\be\label{juayya}
{\bf W}(\bm{\psi},\bm{\psi})={\bf H}(\bm{\psi},\bm{\psi})=1\qquad {\rm for} \qquad \bm{\psi}=|P_1,P_2\rangle\ .
\ee

Finally let us take a closer look at \eqref{AKNSint1}, where $J$ and $T$ are defined by 
\eqref{asdasd1a}.
It is straightforward to see that if we introduce the conjugation
\be\label{JTeq1}
\big[J(u)\big]^{\star}=J(-u^*)\,,\qquad \big[T(u)\big]^{\star}=T(-u^*)\,,
\ee
then the AKNS local IM are Hermitian for any real $k$.
The Hermitian form $\bf{HV}$ that is consistent with the conjugation \eqref{JTeq1}, i.e.,
\be
{\bf HV}\big(\bm{\psi}_2,J(u)\bm{\psi}_1\big)={\bf HV}\big(J(-u^*)\bm{\psi}_2,\bm{\psi}_1\big)\,,\quad
{\bf HV}\big(\bm{\psi}_2,T(u)\bm{\psi}_1\big)={\bf HV}\big(T(-u^*)\bm{\psi}_2,\bm{\psi}_1\big)
\ee
and normalized similar to \eqref{juayya} coincides with $\bf{W}$ and $\bf{H}$ for
$c\ge 2$. However in the domain $c<2$, in full analogy with \eqref{form1as}, the  following relation
holds true
\be
{\bf HV}\big(\bm{\psi}_2,\bm{\psi}_1\big)={\bf W}\big(\bm{\psi}_2,\check{\mathbb{R}}^{({\rm AKNS})}\bm{\psi}_1\big)\ .
\ee

 \section{Conclusion}
Together with the commuting family 
of local IM, the reflection operator(s) are one of the key ingredients
in a variety of integrable structures of Conformal Field Theory.
In this work we used the results of ref.\cite{ET} to compute
the spectrum of the reflection operators in the Fateev integrable
structure. Since the quantum AKNS, KdV and paperclip integrable
structures can be obtained through certain reductions of the Fateev
one, we were able to find the spectrum of their  associated  reflection 
operators as well.
Another result that deserves to be mentioned is the remarkable formula \eqref{ScigLiouveq1} connecting the 
reflection $S$-matrices of the cigar and Liouville CFTs.
\bigskip

%The explicit formulae for the  spectrums turned out rather cumbersome even
%for the quantum KdV integrable structure. In the last case there exists
%an alternative  description of the eigenstates $|{\boldsymbol v}\rangle_P$
%based on the system algebraic equations which is  different than \eqref{jsaysssysa}.
%it would be interesting to express the spectrum of the reflection operator in terms the
%solutions of that system. 

%\bigskip

As was demonstrated in the papers \cite{Bazhanov:2019xvy,Kotousov:2019ygw},  the reflection operators
 are a powerful tool for the study of the scaling limit of the Bethe states
in  integrable spin chains. 
We believe that this is one of the major applications for the results of our work.
In particular, the quantum AKNS integrable structure occurs in the context of
the alternating spin chain \cite{Bazhanov:2019xvy}, while the KdV one is related to the
spin-$\half$ $XXZ$ model \cite{Kotousov:2019ygw}. 
Our result on the spectrum of the paperclip integrable structure
is directly applicable to the study of the critical behaviour of the 
Fateev-Zamolodchikov $\mathbb{Z}_n$-invariant spin chain
(for details see, e.g.,\cite{Bazhanov:2017nzh}). The most general case of the
Fateev integrable structure is expected to occur in the scaling limit of the
higher spin integrable $\mathfrak{sl}(2)$ chains.
\bigskip

We mentioned the construction of the reflection operator for the
quantum Bullough-Dodd integrable structure. Though it has yet to be carried out, 
the calculation of its spectrum 
would be valuable for the study of the scaling behaviour of the
Izergin-Korepin spin chain \cite{Izergin:1980pe}.
Similar to how the quantum KdV integrable structure is obtained from a reduction
of the AKNS one, the Bullough-Dodd  can be derived through
a reduction of the quantum Boussinesq integrable structure \cite{Fioravanti:1995cq,Bazhanov:2001xm}.
The system of algebraic equations 
whose solution sets label the eigenstates for the  Boussinesq integrable structure
was obtained recently in the works \cite{Masoero:2019wqf,BazhanovN}. 
In this case the eigenvalues of the reflection operator are expressed in terms
of the connection coefficients of a certain class of third order ODEs.
For the quantum Getmanov integrable structure
the spectrum of the reflection operator is likewise related to another class
of linear third order differential equations \cite{Lukya}.
Hence the extension of the results of Eremenko and Tarasov to third and higher order
ODEs is of special interest in this regard. Note that the higher order ODEs occur in the
computation of the spectrum of the reflection operator in the
integrable hierarchy for the  Toda theories associated with
the affine Lie (super)algebras \cite{Dorey:2006an,Frenkel:2016gxg,Masoero:2018rel}.
\bigskip

Another application is related to the recent interest 
in constructing the Generalized Gibbs ensemble 
for an integrable QFT \cite{Maloney:2018yrz,Dymarsky:2018iwx}. 
Going beyond the $c=\infty$ limit, 
the  formulation of the Generalized Gibbs ensemble
in all likelihood requires one to properly account for
the reflection operator
and perhaps other non-local IM.

\bigskip

\section*{Acknowledgments}

The authors thank V. Bazhanov,  A.  Eremenko, D. Masoero, V. Tarasov and J. Teschner for stimulating discussions.

\medskip
\noindent
The research of GK is funded by the Deutsche Forschungsgemeinschaft (DFG, German Research
Foundation) under Germany's Excellence Strategy -- EXC 2121 ``Quantum Universe'' -- 390833306.

\medskip
\noindent
This work was done during the second author's visit to the International Institute of Physics at Natal. 
SL is grateful to the IIP for its support and hospitality.

\appendix
\section*{Explicit formulae for ${\Omega^I}_J$, ${\Theta^I}_J$ at the first few levels}

Here we present explicit formulae for the matrices ${\Omega^I}_J$ \eqref{hsysyas} for $L=1,2,3$ 
and ${\Theta^I}_J$ \eqref{Umat1} for $L=1,2$, which are the building blocks
for the Liouville \eqref{jasusaus} and cigar \eqref{aoasisaisa} reflection $S$-matrices.
\bigskip

The oscillator basis for the Fock space ${\cal F}_P$ 
reads as
\be
\bm{e}_{(1)}=a_{-1}|P\rangle\nonumber
\ee
for level one;
\be
\bm{e}_{(1,1)}=a_{-1}^2|P\rangle\,,\qquad \bm{e}_{(2)}=a_{-2}|P\rangle\nonumber
\ee
for the case $L=2$; and
\be
\bm{e}_{(1,1,1)}=a_{-1}^3|P\rangle\,,\qquad
\bm{e}_{(1,2)}=a_{-1} a_{-2}|P\rangle\,,\qquad \bm{e}_{(3)}=a_{-3}|P\rangle\nonumber
\ee
for the third level. As was mentioned in the main body of the text,
the Fock space admits the structure of the Verma module for
the Virasoro algebra, with $|P\rangle$ being the highest weight vector
with the conformal dimension $\Delta=P^2+\tfrac{1}{4}\,Q^2$ and the central charge
$c=1+6Q^2$. The Virasoro basis is obtained by acting
with the $L_{-n}$ with $n>0$ on $|P\rangle$.
Using  eqs.\,\eqref{ososossi} that express the Virasoro generators
in terms of the Heisenberg ones, this basis can be re-written in terms
of the $\bm{e}_I$ presented above. A straightforward calculation
gives
\be\label{VirLvl1}
\bm{t}_{(1)}\equiv L_{-1}|P\rangle=(2P-\ri Q)\bm{e}_{(1)}\,,\nonumber
\ee
while
\be\label{VirLvl2}
\bm{t}_{(1,1)}\equiv L_{-1}^2|P\rangle=(2P-\ri Q)^2\bm{e}_{(1,1)}+(2P-\ri Q)\bm{e}_{(2)}\,,\quad
\bm{t}_{(2)}\equiv L_{-2}|P\rangle=\bm{e}_{(1,1)}+2(P-\ri Q)\bm{e}_{(2)}\ .\nonumber
\ee
At the third level one has
\bea\label{VirLvl3}
\bm{t}_{(1,1,1)}\equiv L_{-1}^3|P\rangle&=& (2P-\ri Q)^3\,\bm{e}_{(1,1,1)}+3(2P-\ri Q)^2\,\bm{e}_{(1,2)}
+2(2P-\ri Q)\,\bm{e}_{(3)}\nonumber \\[0.2cm]
\bm{t}_{(1,2)}\equiv L_{-1}L_{-2}|P\rangle &=&(2P-\ri Q)\,\bm{e}_{(1,1,1)}+
2\big(1+(P-\ri Q)(2P-\ri Q)\big)\,\bm{e}_{(1,2)}+
4(P-\ri Q)\,\bm{e}_{(3)}\nonumber \\[0.2cm]
\bm{t}_{(3)}\equiv L_{-3}|P\rangle &=&
2\,\bm{e}_{(1,2)}+(2P-3\ri Q)\,\bm{e}_{(3)}\ .\nonumber
\eea
The matrix elements ${\Omega^I}_J$ \eqref{hsysyas} are easily read off from
the above formulae. For level $L=1$:
\be
{\Omega^{(1)}}_{(1)}=2P-\ri Q\nonumber\ .
\ee
At the second level
\be
\bm{\Omega}=\left(\begin{array}{cc}
{\Omega^{(1,1)}}_{(1,1)} & {\Omega^{(1,1)}}_{(2)} \\[0.2cm]
{\Omega^{(2)}}_{(1,1)} & {\Omega^{(2)}}_{(2)}
\end{array}\right)=\left(\begin{array}{cc}
(2P-\ri Q)^2 & 1 \\[0.2cm]
2P-\ri Q & 2(P-\ri Q)
\end{array}\right)\ ,\nonumber
\ee
while at the third level, using similar matrix notation,
\be
\bm{\Omega}=\left(\begin{array}{ccc}
(2P-\ri Q)^3 & 2P-\ri Q & 0 \\[0.2cm]
3(2P-\ri Q)^2 & 2+2(P-\ri Q)(2P-\ri Q)& 2 \\[0.2cm]
2(2P-\ri Q) & 4(P-\ri Q) & 2P-3\ri Q
\end{array}\right)\nonumber\ .
\ee
\bigskip

The computation for ${\Theta^I}_J$ \eqref{Umat1} is completely
analogous to that of ${\Omega^I}_J$ described above.
In this case one needs to relate the basis \eqref{Wbasis10} built from the Fourier modes
of the currents $W_2(u)$ and $W_3(u)$ with the one obtained by acting with the
Heisenberg modes $a_{-m}$, $b_{-m}$ $(m>0)$ on the Fock vacuum $|P_1,P_2\rangle$,
see eq.\,\eqref{Obasis10}. A straightforward calculation using eqs.\,\eqref{Weq1},\,\eqref{Wmodes1}
and \eqref{asympeq1} allows one to express the $W$-modes in terms of the 
Heisenberg generators. The result reads as
\bea
\widetilde{W}_2(m)&=&\sum_{j=-\infty}^\infty\big(a_j\,a_{m-j}+b_j\,b_{m-j}\big)+\ri m\, Q\ a_m
\ \ \ \ \ \ \ \ \qquad (m\ne 0)\nonumber\\[0.2cm]
\widetilde{W}_3(m)&=&\frac{4Q^2+6}{3}\!\!\sum_{i+j+l=m}:\! b_i b_j b_{l}\!:
+\,2\!\!\sum_{i+j+l=m}:\! a_i a_j\!: b_{l}
-\frac{\ri}{Q}\,\sum_{l=-\infty}^\infty l\,\big(a_l\, b_{m-l}-(1+2Q^2)\,b_l \,a_{m-l}\big)\nonumber \\[0.2cm]
&-&\frac{2+Q^2+m^2(1+2Q^2)}{6}\ b_m\, \qquad\ \ \ \ \ \ \ \ \, \ \ \ \ \ \ \qquad (m\ne 0)\ .\nonumber
\eea
Here $Q=-\frac{1}{\sqrt{k}}$ and
the monomials occurring in the sums are normal ordered so that
the creation operators are always placed to the left
of the annihilation operators.
At the first level, the basis vectors
\be
\bm{v}_{(1)}=\widetilde{W}_2(-1)|P_1,P_2\rangle,\,
\qquad \bm{v}_{(1')}=\widetilde{W}_3(-1)|P_1,P_2\rangle\nonumber
\ee
are expressed in terms of the Heisenberg 
basis:
\be
\bm{e}_{(1)}={a}_{-1}|P_1,P_2\rangle,\,
\qquad \bm{e}_{(1')}={b}_{-1}|P_1,P_2\rangle\nonumber
\ee
 as 
\bea
\bm{v}_{(1)}&=&(2P_1-\ri\,Q)\,\bm{e}_{(1)}
+2P_2\,\bm{e}_{(1')}\nonumber \\[0.3cm]
\bm{v}_{(1')}&=&P_2\,\Big(4P_1+\frac{\ri}{Q}\,\Big)\,\bm{e}_{(1)}+
B\,
\bm{e}_{(1')}\,,\nonumber
\eea
where
\be
B=
2(2Q^2+3)P_2^2+P_1\big(2P_1-\ri(2Q+Q^{-1})\big)-\frac{1}{2}\,(Q^2+1)\nonumber\ .
\ee
Hence the matrix elements ${\Theta^I}_J$ \eqref{Umat1} at level $L=1$ are given by
\be
\bm{\Theta}=\left(\begin{array}{cc}
{\Theta^{(1)}}_{(1)} & {\Theta^{(1)}}_{(1')} \\[0.2cm]
{\Theta^{(1')}}_{(1)} & {\Theta^{(1')}}_{(1')}
\end{array}\right)=\left(\begin{array}{cc}
2P_1-\ri Q & P_2\,\big(4P_1+\frac{\ri}{Q}\big) \\[0.2cm]
2P_2 & B
\end{array}\right)\ .\nonumber
\ee

The level $L=2$ subspace of the Fock space ${\cal F}_{P_1,P_2}$ is five dimensional.
It is spanned by the states
\be
\bm{e}_{(1,1)}=a_{-1}^2|P_1,P_2\rangle,\,\qquad \bm{e}_{(2)}=a_{-2}|P_1,P_2\rangle\,,\qquad
\bm{e}_{(1',1')}=b_{-1}^2|P_1,P_2\rangle\,,\nonumber
\ee
\vskip-0.2cm
\be
\bm{e}_{(2')}=b_{-2}|P_1,P_2\rangle\,,\qquad \bm{e}_{(1,1')}=a_{-1}b_{-1}|P_1,P_2\rangle\ .\nonumber
\ee
There are five vectors of the form \eqref{Wbasis10} belonging to this subspace:
\be
\bm{v}_{(1,1)}=\widetilde{W}_2^2(-1)|P_1,P_2\rangle,\,\qquad \bm{v}_{(2)}=\widetilde{W}_2(-2)|P_1,P_2\rangle\,,\qquad
\bm{v}_{(1',1')}=\widetilde{W}_3^2(-1)|P_1,P_2\rangle\,,\nonumber
\ee
\be
\bm{v}_{(2')}=\widetilde{W}_3(-2)|P_1,P_2\rangle\,,\qquad 
\bm{v}_{(1,1')}=\widetilde{W}_2(-1)\widetilde{W}_3(-1)|P_1,P_2\rangle\ .\nonumber
\ee
In terms of the Heisenberg basis, they are given by
\bea
\bm{v}_{(1,1)}&=&(2P_1-\ri Q)^2\,\bm{e}_{(1,1)}+(2P_1-\ri Q)\,\bm{e}_{(2)}+4P_2^2\,\bm{e}_{(1',1')}
+2P_2\,\bm{e}_{(2')}+4P_2(2P_1-\ri Q)\,\bm{e}_{(1,1')}
\nonumber\\[0.15cm]
\bm{v}_{(2)}&=&\bm{e}_{(1,1)}+2(P_1-\ri Q)\,\bm{e}_{(2)}+\bm{e}_{(1',1')}+2P_2\,\bm{e}_{(2')}\nonumber\\[0.15cm]
\bm{v}_{(2')}&=&2P_2\,\bm{e}_{(1,1)}-\frac{2\ri P_2}{Q}\,(2\ri P_1 Q-1)\bm{e}_{(2)}+2P_2(2Q^2+3)\,\bm{e}_{(1',1')}
+\frac{1}{2Q}\big[4Q\,P_1^2\nonumber \\[0.15cm]
&+&4Q(2Q^2+3)\,P_2^2-4\ri (2Q^2+1)\,P_1-Q(3Q^2+2)\,\big]\,
\bm{e}_{(2')}+2(2P_1-\ri Q)\,\bm{e}_{(1,1')}\nonumber\\[0.15cm]
\bm{v}_{(1,1')}&=&-\frac{\ri P_2}{Q}\,(2P_1-\ri Q)(4\ri Q\,P_1-1)\,\bm{e}_{(1,1)}
-\frac{\ri P_2}{Q}\,(4\ri Q P_1-1)\,\bm{e}_{(2)}\nonumber \\[0.15cm]
&+&\frac{P_2}{Q}\,\big[4Q\,P_1^2+4Q(2Q^2+3)\,P_2^2-2\ri(2Q^2+1)\,P_1-Q(Q^2+1)
\big]\bm{e}_{(1',1')}\nonumber \\[0.15cm]
&+&\frac{1}{2Q}\,\big[4Q\,P_1^2+4Q\,(2Q^2+3)\,P_2^2
-2\ri(2Q^2+1)\,P_1-Q(Q^2+1)\big]\,\bm{e}_{(2')}\nonumber \\[0.15cm]
&+&\frac{1}{2Q}\,
\big[8Q\,P_1^3+8Q\,(2Q^2+5)\,P_1 P_2^2-4\ri(3Q^2+1)\,P_1^2-4\ri(2Q^4+3Q^2-1)\,P_2^2\nonumber \\[0.15cm]
&-&
2Q(3Q^2+2)\,P_1+\ri Q^2(Q^2+1)\big]\,\bm{e}_{(1,1')}
\nonumber
\eea
The expression for $\bm{v}_{(1',1')}$ is rather cumbersome and reads as
\bea
\bm{v}_{(1',1')}&=&\frac{1}{2Q^2}\,\big[32Q^2\,P_1^2 P_2^2+16\ri Q\,P_1P_2^2+4Q^2\,P_1^2
+2(4Q^4+6Q^2-1)\,P_2^2-2\ri Q(1+2Q^2)P_1\nonumber \\[0.2cm]
&-&Q^2(1+Q^2)\big]\,\bm{e}_{(1,1)}+
\frac{1}{4Q^2}\big[16Q^2\,P_1^3+16Q^2(2Q^2+5)P_1\,P_2^2-4\ri Q(2Q^2-1)\,P_1^2\nonumber \\[0.2cm]
&+&4\ri Q\,(4Q^4+12Q^2+11)\,P_2^2
+2\,(2Q^4+6Q^2+3)\,P_1-\ri Q(Q^2+1)(2Q^2+3)\big]\,\bm{e}_{(2)}\nonumber \\[0.2cm]
&+&
\frac{1}{4Q^2}\big[16Q^2\,P_1^4+16Q^2(2Q^2+3)^2\,P_2^4+32Q^2(2Q^2+3)\,P_1^2 P_2^2
-16\ri Q(2Q^2+1)\,P_1^3\nonumber \\[0.2cm]
&-&16\ri Q(2Q^2+1)(2Q^2+3)\,P_1 P_2^2
-4(2Q^4+1)P_1^2+8\,Q^2\,(Q^2+2)(2Q^2+3)\,P_2^2\nonumber \\[0.2cm]
&-&4\ri Q(Q^2+2)(2Q^2+1)\,P_1
-Q^2\,(5+8Q^2+3Q^4)\big]\,\bm{e}_{(1',1')}+\frac{P_2}{2Q^2}\,
\big[8Q^2(2Q^2+5)\,P_1^2\nonumber \\[0.2cm]
&+&8Q^2(2Q^2+3)^2\,P_2^2
-4\ri Q(4Q^4+12Q^2+5)\,P_1-(4Q^6+10Q^4+2Q^2-3)\big]\,\bm{e}_{(2')}
\nonumber \\[0.2cm]
&+&\frac{P_2}{Q^2}\big[16Q^2\,P_1^3+16Q^2(2Q^2+3)\,P_1P_2^2-4\ri Q(4Q^2+1)\,P_1^2+
4\ri Q(2Q^2+3)\,P_2^2\nonumber \\[0.2cm]
&-&2
(2Q^4-4Q^2-1)\,P_1-\ri Q(Q^2-1)\big]\bm{e}_{(1,1')}\ .
\nonumber
\eea
The matrix elements ${\Theta^I}_J$  \eqref{Umat1}  at the second level
follow immediately from the above formulae.

\end{document}